\documentclass[a4paper,english]{aa}
\usepackage{times}
\usepackage[T1]{fontenc}
\usepackage[latin1]{inputenc}
\usepackage{graphicx}
\usepackage[authoryear]{natbib}

\makeatletter

\newcommand{\noun}[1]{\textsc{#1}}
\providecommand{\tabularnewline}{\\}

\usepackage{txfonts}
\usepackage{aalongtable}
\bibpunct{(}{)}{;}{a}{}{,}

\usepackage{babel}
\makeatother
\begin{document}

\title{Spectrophotometric properties of galaxies at intermediate redshifts
(z\textasciitilde{}0.2--1.0)\thanks{Based on observations collected at the Very Large Telescope, European Southern Observatory, Paranal, Chile (ESO Programs 64.O-0439, 65.O-0367, 67.B-0255, 69.A-0358, and 72.A-0603)}}

\subtitle{II. The Luminosity -- Metallicity relation\thanks{Tables~\ref{ohdata} and~\ref{ohdatacl} are only available in electronic form at the CDS via anonymous ftp to cdsarc.u-strasbg.fr (130.79.128.5) or via http://cdsweb.u-strasbg.fr/cgi-bin/qcat?J/A+A/}}

\author{F. Lamareille\inst{1} \and T. Contini\inst{1} \and J. Brinchmann\inst{2,3} \and J.-F.
Le Borgne\inst{1} \and S. Charlot\inst{2,4} \and J. Richard\inst{1} }

\institute{Laboratoire d'Astrophysique de Toulouse et Tarbes (LATT - UMR 5572),
Observatoire Midi-Pyrénées, 14 avenue E. Belin, F-31400 Toulouse,
France\and Max-Planck Institut für Astrophysik, Karl-Schwarzschild-Strasse
1 Postfach 1317, D-85741 Garching, Germany\and Centro de Astrof\'isica
da Universidade do Porto, Rua das Estrelas - 4150-762 Porto, Portugal\and Institut
d'Astrophysique de Paris, CNRS, 98 bis Boulevard Arago, F-75014 Paris,
France}

\offprints{F. Lamareille, e-mail: \texttt{flamare@ast.obs-mip.fr}}

\date{Received ; Accepted}

\abstract{We present the gas-phase oxygen abundance (O/H) for a sample of 131
star-forming galaxies at intermediate redshifts ($0.2<z<1.0$). The
sample selection, the spectroscopic observations (mainly with VLT/FORS)
and associated data reduction, the photometric properties, the emission-line
measurements, and the spectral classification are fully described
in a companion paper (Paper~I). We use two methods to estimate the
O/H abundance ratio: the {}``standard'' $R_{\mathrm{23}}$ method
which is based on empirical calibrations, and the CL01 method which
is based on grids of photo-ionization models and on the fitting of
emission lines. For most galaxies, we have been able to solve the
problem of the metallicity degeneracy between the high- and low-metallicity
branches of the O/H vs. $R_{\mathrm{23}}$ relationship using various
secondary indicators. The luminosity -- metallicity ($L-Z$) relation
has been derived in the $B$- and $R$-bands, with metallicities derived
with the two methods ($R_{\mathrm{23}}$ and CL01). In the analysis,
we first consider our sample alone and then a larger one which includes
other samples of intermediate-redshift galaxies drawn from the literature.
The derived $L-Z$ relations at intermediate redshifts are very similar
(same slope) to the $L-Z$ relation obtained for the local universe.
Our sample alone only shows a small, not significant, evolution of
the $L-Z$ relation with redshift up to $z\sim1.0$. We only find
statistical variations consistent with the uncertainty in the derived
parameters. Including other samples of intermediate-redshift galaxies,
we find however that galaxies at $z\sim1$ appear to be metal-deficient
by a factor of $\sim3$ compared with galaxies in the local universe.
For a given luminosity, they contain on average about one third of
the metals locked in local galaxies. }

\maketitle

\section{Introduction}

The understanding of galaxy formation and evolution has entered a
new era since the advent of 10-m class telescopes and the associated
powerful multi-object spectrograph, such as VIMOS on the VLT or DEIMOS
on Keck. It is now possible to collect spectrophotometric data for
large samples of galaxies at various redshifts, in order to compare
the physical properties (star formation rate, extinction, metallicity,
etc) of galaxies at different epochs of the Universe, using the results
of recent surveys such as the {}``Sloan Digital Sky Survey'' (SDSS,
\citealt{Abazajian:2003AJ....126.2081A,Abazajian:2004AJ....128..502A}
and the {}``2 degree Field Galaxy Redshift Survey'' (2dFGRS, \citealt{Colless:2001MNRAS.328.1039C})
as references in the local Universe.

The correlation between galaxy metallicity and luminosity in the local
universe is one of the most significant observational results in galaxy
evolution studies. \citet{Lequeux:1979A&A....80..155L} first revealed
that the oxygen abundance O/H increases with the total mass of irregular
galaxies. To avoid several problems in the estimate of dynamical masses
of galaxies, especially for irregulars, absolute magnitudes are commonly
used. The luminosity -- metallicity ($L-Z$) relation for irregulars
was later confirmed by \citet*{Skillman:1989ApJ...347..875S}, \citet{Richer:1995ApJ...445..642R}
and \citet{Pilyugin:2001A&A...374..412P} among others. Subsequent
studies have extended the $L-Z$ relation to spiral galaxies \citep{Garnett:1987ApJ...317...82G,Zaritsky:1994ApJ...420...87Z,Garnett:1997ApJ...489...63G,Pilyugin:2000A&A...358...72P},
and to elliptical galaxies \citep{Brodie:1991ApJ...379..157B}. The
luminosity correlates with metallicity over $\sim10$ magnitudes in
luminosity and 2 dex in metallicity, with indications that the relationship
may be environmental- \citep{Vilchez:1995AJ....110.1090V} and morphology-
\citep{Mateo:1998ARA&A..36..435M} free. This suggests that similar
phenomena govern the $L-Z$ over the whole Hubble sequence, from irregular/spirals
to ellipticals (e.g. \citealp*{Garnett:2002ApJ...581.1019G,Pilyugin:2004A&A...425..849P}).
Recently, the $L-Z$ relation in the local universe has been derived
using the largest datasets available so far, namely the 2dFGRS \citep{Lamareille:2004MNRAS.350..396L}
and the SDSS \citep{Tremonti:2004astro.ph..5537T}. The main goal
of this paper is to derive the $L-Z$ relation for a sample of intermediate-redshift
star-forming galaxies, and investigate how it compares with the local
relation.

Recent studies of the $L-Z$ relation at intermediate redshifts have
provided conflicting evidence for any change in the relation and no
consensus has been reached. \citet{Kobulnicky:1999ApJ...511..118K}
and \citet*{Lilly:2003ApJ...597..730L} found their samples of intermediate-redshift
galaxies to conform to the local $L-Z$ relation without any significant
evolution of this relation out to $z\sim1$. In contrast, other authors
(e.g. \citealp{Kobulnicky:2003ApJ...599.1006K,Maier:2004A&A...418..475M,Liang:2004A&A...423..867L,Hammer:2004astroph0410518,Kobulnicky:2004ApJ...617..240K})
have recently claimed that both the slope and zero point of the $L-Z$
relation evolve with redshift, the slope becoming steeper and the
zero point decreasing at early cosmic time. This would mean that galaxies
of a given luminosity are more metal-poor at higher redshift, showing
a decrease in average oxygen abundance by $\sim0.15$ dex from $z=0$
to $z=1$.

In this paper, we present gas-phase oxygen abundance measurements
for 131 star-forming galaxies in the redshift range $0.2<z<1.0$.
The sample selection, the spectroscopic observations and data reduction,
the photometric properties, the emission-line measurements, and the
spectral classification of a sample of 141 intermediate-redshift emission-line
galaxies are detailed in \citet{Lamareille:2005AA}, hereafter Paper~I.
Among the sample of 131 star-forming galaxies, 16 objects may contain
a contribution from a low-luminosity active galactic nucleus and are
thus flagged as {}``candidate'' star-forming galaxies (see Paper~I
for details). Spectra were acquired mainly with the FORS1/2 (FOcal
Reducer Spectrograph) instrument on the VLT, with the addition of
LRIS (Low Resolution Imaging Spectrograph) observations on the Keck
telescope, and galaxies selected from the {}``Gemini Deep Deep Survey''
(GDDS, \citealt{Abraham:2004AJ....127.2455A}) public data release.
The spectra are sorted into three sub-samples with different selection
criteria: the {}``CFRS sub-sample'' contains emission-line galaxies
selected from the {}``Canada-France Redshift Survey'' \citep{Lilly:1995ApJ...455...50L},
the {}``CLUST sub-sample'' contains field galaxies randomly selected
behind lensing clusters, and the {}``GDDS sub-sample'' stands for
galaxies taken from the GDDS survey.

These new data increase significantly the number of metallicity estimates
available for this redshift range and are among the highest quality
spectra yet available for the chemical analysis of intermediate-redshift
galaxies. In addition to providing new constraints on the chemical
enrichment of galaxies over the last $\sim8$ Gyr, we hope that these
measurements will be useful in modeling the evolution of galaxies
on cosmological timescales. These new data are combined with existing
emission-line measurements from the literature to assess the chemical
evolution of star-forming galaxies out to $z=1$. In the near future,
this work will be extended to samples of thousands of galaxies up
to $z\sim1.5$, thanks to the massive ongoing deep spectroscopic surveys
such as the {}``VIMOS VLT Deep Survey'' (VVDS, \citealt{LeFevre:2004A&A...417..839L}).

The paper is organized as follows: Sect.~\ref{sec:oxyabun} discusses
the methods we have used to estimate the gas-phase oxygen abundance,
and how we have addressed the issue of degeneracy in the determination
of the oxygen abundance from strong emission lines. In Sect.~\ref{sec:lz}
we study the relation between the luminosities and the metallicities
of our sample galaxies, combined with other samples of intermediate-redshift
galaxies published so far. Finally, in Sect.~\ref{sec:evolz}, we
discuss the possible evolution of the $L-Z$ relation with redshift.

Throughout this paper, we use the WMAP cosmology \citep{Spergel:2003ApJS..148..175S}:
$H_{\mathrm{0}}=71$ km s$^{-1}$ Mpc$^{-1}$, $\Omega_{\Lambda}=0.73$
and $\Omega_{\mathrm{\mathrm{m}}}=0.27$. The magnitudes are given
in the AB system.

\section{Gas-phase oxygen abundance\label{sec:oxyabun}}

Emission lines are the primary source of information regarding gas-phase
chemical abundances within star-forming regions. The {}``direct''
method for determining the chemical composition requires the electron
temperature and the density of the emitting gas (e.g. \citealp{Osterbrock:1989agna.book.....O}).
Unfortunately, a direct heavy-element abundance determination, based
on measurements of the electron temperature and density, cannot be
obtained for faint galaxies. The {[}O\noun{iii}{]}$\lambda$4363
auroral line, which is the most commonly applied temperature indicator
in extragalactic H\textsc{ii} regions, is typically very weak and
rapidly decreases in strength with increasing abundance; it is expected
to be of order $10^{2}-10^{3}$ times fainter than the {[}O\noun{iii}{]}$\lambda$5007
line.

\begin{figure}
\begin{center}\includegraphics[%
  width=1.0\columnwidth]{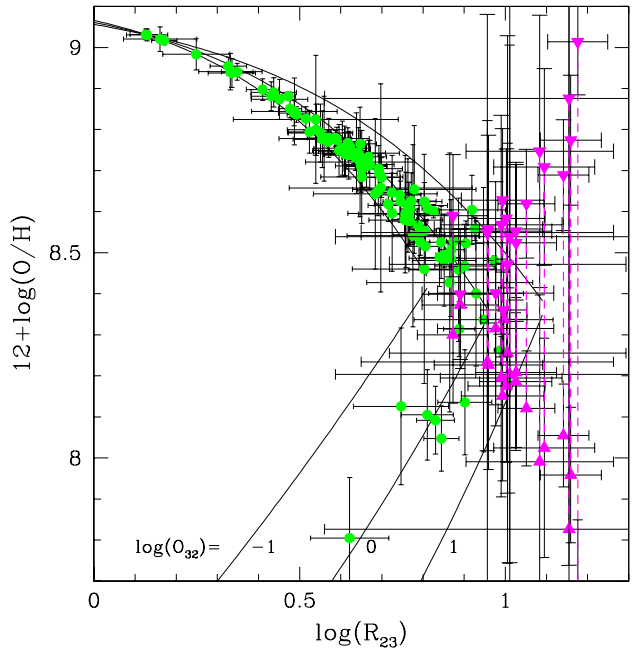}\end{center}

\caption{Calibration of gas-phase oxygen abundance as a function of the strong-line
ratio log($R_{\mathrm{23}}$) (based on equivalent width measurements).
Calibration curves from \citet{McGaugh:1991ApJ...380..140M} (analytical
formulae from \citealp{Kobulnicky:1999ApJ...514..544K}) are shown
for three different values of the ionization parameter expressed in
terms of the observable line ratio log($O_{\mathrm{32}}$) ($-$1,
0 and 1). Our sample of intermediate-redshift galaxies are plotted
with the following symbols: filled circles are for {}``normal''
objects, while the triangles represent intermediate-metallicity galaxies
(see Sect.~\ref{sub:ohResults} for details).}

\label{ohcalib}
\end{figure}

Given the absence of reliable {[}O\noun{iii}{]}$\lambda$4363 detections
in our spectra of faint objects, alternative methods for deriving
nebular abundances must be employed that rely on observations of the
bright lines alone. Empirical methods to derive the oxygen abundance
exploit the relationship between O/H and the intensity of the strong
lines via the parameter $R_{\mathrm{23}}$=({[}O\noun{iii}{]}$\lambda\lambda$4959+5007+{[}O\noun{ii}{]}$\lambda$3727)/H$\beta$
(see Fig.~\ref{ohcalib}).

Many authors have developed techniques for converting $R_{\mathrm{23}}$
into oxygen abundance, both for the metal-poor \citep*{Pagel:1980MNRAS.193..219P,Skillman:1989ApJ...347..883S,Pilyugin:2000A&A...362..325P}
and metal-rich \citep{Pagel:1979MNRAS.189...95P,Edmunds:1984MNRAS.211..507E,McCall:1985ApJS...57....1M}
regimes. On the \emph{upper}, metal-rich branch of the $R_{\mathrm{23}}$
vs. O/H relationship, $R_{\mathrm{23}}$ increases as metallicity
decreases via reduced cooling, elevated electronic temperatures, and
a higher degree of collisional excitation. However, the relation between
$R_{\mathrm{23}}$ and O/H becomes degenerate below 12+log(O/H)$\sim8.4$
($Z\sim0.3\, Z_{\odot}$) and $R_{\mathrm{23}}$ reaches a maximum
(see Fig.~\ref{ohcalib}).

For oxygen abundances below 12+log(O/H)$\sim8.2$, $R_{\mathrm{23}}$
decreases with decreasing O/H --- this defines the \emph{lower}, metal-poor
branch. The decrease in $R_{\mathrm{23}}$ takes place because the
greatly reduced oxygen abundance offsets the effect of reduced cooling
and raised electron temperatures caused by the lower metal abundance.
In this regime, the ionization parameter, defined by the emission-line
ratio $O_{\mathrm{32}}$={[}O\noun{iii}{]}$\lambda\lambda$4959+5007/{[}O\noun{ii}{]}$\lambda$3727,
also becomes important \citep{McGaugh:1991ApJ...380..140M}.

The typical spread in the $R_{\mathrm{23}}$ vs. O/H relationship
is $\pm0.15$ dex, with a slightly larger spread ($\pm0.25$ dex)
in the turnaround region near 12+log(O/H)$\sim8.4$. This dispersion
reflects the uncertainties in the calibration of the $R_{\mathrm{23}}$
method which is based on photo-ionization models and observed H\noun{ii}
regions. However, the most significant uncertainty involves deciding
whether an object lies on the \emph{upper}, metal-rich branch, or
on the \emph{lower}, metal-poor branch of the curve (see Fig.~\ref{ohcalib}).

In this paper, we use the strong-line method to estimate the gas-phase
oxygen abundance of our sample galaxies using the emission line measurements
reported in Paper I. Two different methods are considered: the $R_{\mathrm{23}}$
method \citep{McGaugh:1991ApJ...380..140M,Kewley:2002ApJS..142...35K}
which is based on empirical calibrations between oxygen-to-hydrogen
emission-line ratios and the gas-phase oxygen abundance, and the \citet[hereafter CL01]{Charlot:2001MNRAS.323..887C}
method which is based on the simultaneous fit of the luminosities
of several emission lines using a large grid of photo-ionization models.
The CL01 method has the potential of breaking the degeneracies in
the determination of oxygen abundance (see CL01 for details), but
we must point out that both the $R_{\mathrm{23}}$ estimator and the
CL01 approach are limited by the O/H degeneracy when only a few emission
lines are used, unless further information exists. However, the use
of these two methods provides an important consistency check.

\subsection{The $R_{\mathrm{23}}$ method\label{sub:Using-empirical-calibrations}}

We first discuss the commonly used $R_{\mathrm{23}}$ method which
is based on empirical calibrations as mentioned above. This method
uses two emission-line ratios: $R_{\mathrm{23}}$=({[}O\noun{iii}{]}$\lambda\lambda$4959+5007+{[}O\noun{ii}{]}$\lambda$3727)/H$\beta$,
and $O_{\mathrm{32}}$={[}O\noun{iii}{]}$\lambda\lambda$4959+5007/{[}O\noun{ii}{]}$\lambda$3727.
Analytical expressions between the gas-phase oxygen-to-hydrogen abundance
ratio and these two emission-line ratios are found in \citet*{Kobulnicky:1999ApJ...514..544K},
both for the metal-poor \emph{(}lower) and metal-rich \emph{(}upper)
branches (see Fig.~\ref{ohcalib}).

\citet{Kobulnicky:2003ApJ...599.1031K} have shown that equivalent
widths can be used in the $R_{\mathrm{23}}$ method instead of line
fluxes. We will take advantage of this here since this gives equivalent
results in the $R_{\mathrm{23}}$ method. This is very useful for
objects with no reddening estimate as no reddening correction has
to be applied on equivalent-width measurements, assuming that the
attenuation in the continuum and emission lines is the same (see Paper~I
for a comparison between the equivalent width and dust-corrected flux
{[}O\noun{ii}{]}$\lambda$3727/H$\beta$ ratios). This method has
already been applied by \citet{Kobulnicky:2004ApJ...617..240K} on
intermediate-redshift galaxies.

To be more confident we have compared the gas-phase oxygen abundances
we found with equivalent widths or with dust-corrected fluxes on the
$24$ galaxies where a correction for dust reddening was possible
(i.e. H$\alpha$ and H$\beta$ emission lines observed). We find very
good agreement with no bias and the rms of the residuals around the
$y=x$ line is $0.1$ dex only.

\subsubsection{Breaking the double-value degeneracy in the $R_{\mathrm{23}}$ method\label{sub:Breaking-the-degeneracy}}

We explore here different methods to break the degeneracy in the determination
of O/H with the $R_{\mathrm{23}}$ method, which leads to two possible
values of the gas-phase oxygen abundance for a given $R_{\mathrm{23}}$
line ratio (low- and high-metallicity, as discussed in Sect~\ref{sec:oxyabun}).

A number of alternative abundance indicators have been used in previous
works to break this degeneracy, e.g. {[}N\noun{ii}{]}$\lambda$6584/{[}O\noun{iii}{]}$\lambda$5007
\citep{Alloin:1979A&A....78..200A}, {[}N\noun{ii}{]}$\lambda$6584/{[}O\noun{ii}{]}$\lambda$3727
\citep{McGaugh:1994ApJ...426..135M}, {[}N\noun{ii}{]}$\lambda$6584/H$\alpha$
\citep{VanZee:1998AJ....116.2805V}, and galaxy luminosity \citep*{Kobulnicky:1999ApJ...514..544K}.

The most commonly used prescription uses the N2={[}N\noun{ii}{]}$\lambda$6584/H$\alpha$
line ratio as a secondary, non-degenerate, indicator of the metallicity
\citep{VanZee:1998AJ....116.2805V}. We thus discuss the results we
get with this method and then introduce a new method we call the $L$
diagnostic.

\paragraph{The N2 diagnostic.}

\begin{figure}
\begin{center}\includegraphics[%
  width=1.0\columnwidth,
  keepaspectratio]{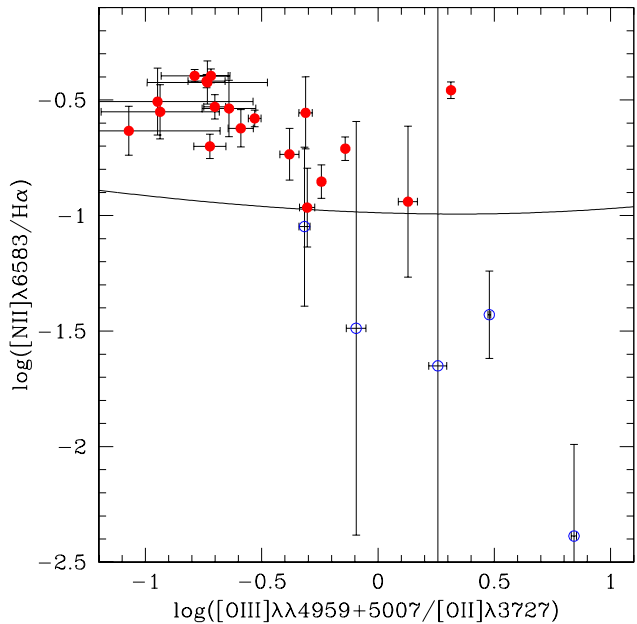}\end{center}

\caption{The N2 separation between high- and low-metallicity galaxies as a
function of $\log(O_{\mathrm{32}})$. The solid line shows the theoretical
limit between low- and high-metallicity galaxies, converted into a
value of log({[}N\noun{ii}{]}$\lambda$6584/H$\alpha$) using the
relation of \citet{VanZee:1998AJ....116.2805V} (see text for details).
The filled circles are high-metallicity objects and open circles low-metallicity
ones for our sample of 25 spectra for which this diagnostic was applied.}

\label{ohlim}
\end{figure}

We want to check if the N2 diagnostic can be used on our data independently
of the ionization parameter. Figure~\ref{ohlim} shows the theoretical
limit between low- and high-metallicity regimes as a function of $\log(O_{\mathrm{32}})$
which is calculated with the following procedure: \emph{(i)} for each
value of $\log(O_{\mathrm{32}})$ we take the maximum value of $\log(R_{\mathrm{23}})$
(i.e. at the turnaround point of the O/H vs. $R_{\mathrm{23}}$ relationship,
see Fig.~\ref{ohcalib}), \emph{(ii)} we calculate the associated
gas-phase oxygen abundance, which is the separation between low- and
high- metallicity regimes, \emph{(iii)} we calculate the associated
value of log({[}N\noun{ii}{]}$\lambda$6584/H$\alpha$) (equation
(1) of \citealp{VanZee:1998AJ....116.2805V}). Our data sample is
also shown in Fig.~\ref{ohlim}. To first order, the observed limit
between low- and high-metallicity objects seems independent of $\log(O_{\mathrm{32}})$.
We thus adopt log({[}N\noun{ii}{]}$\lambda$6584/H$\alpha$)$>-1$,
as our criterion for placing galaxies on the high-metallicity branch
--- in agreement with the standard limit used in the literature \citep{Contini:2002MNRAS.330...75C}.
We note that low-metallicity galaxies show big error bars for the
N2 index. This is explained by a weak {[}N\noun{ii}{]}$\lambda$6584
emission line in this regime.

The emission lines required for the N2 method can only be seen in
our optical spectra for the lowest-redshift galaxies. For the 24 spectra
to which this diagnostic was applied, we find 5 (21\%) low-metallicity
galaxies and 19 (79\%) metal-rich galaxies.

\paragraph{The \(L\) diagnostic.\label{sub:The-L-Z-diagnostic}}

The N2 indicator needs the {[}N\noun{ii}{]}$\lambda$6584 and H$\alpha$
emission lines which are not in the wavelength range of most of our
spectra because of their high redshift. Although the low-redshift
sample shows a high fraction of high-metallicity objects, it is clear
that we cannot make this assumption in general. Thus we need to find
another way to break the O/H degeneracy using information from the
blue part of the spectrum only. One possibility is to break the degeneracy
by creating an hybrid method including additional physical parameters
for the galaxies such as the intensity of the 4000 \AA{} break, the
$u-r$ color or the intensity of the blue Balmer emission-lines. We
do not, however, find any clear correlation between these parameters
and the metallicities in our sample of 24 galaxies with the N2 diagnostic
available.

\begin{figure}
\begin{center}\includegraphics[%
  width=1.0\columnwidth,
  keepaspectratio]{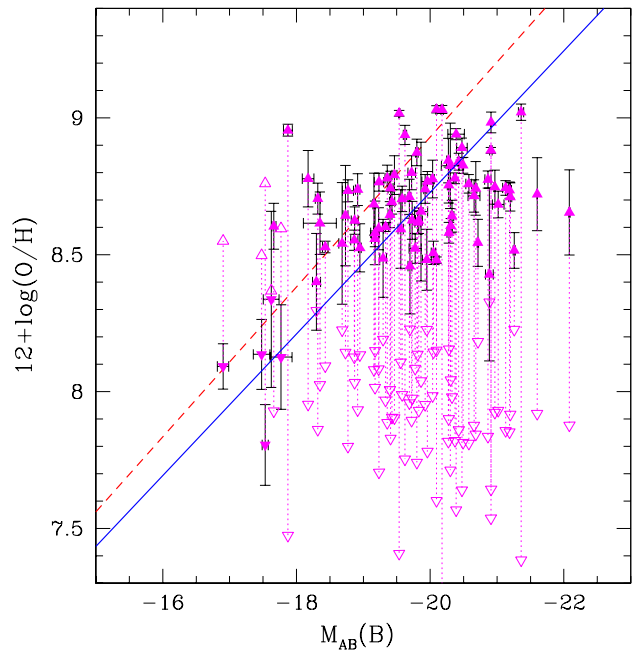}\end{center}

\caption{The $L$ diagnostic for breaking the degeneracy in the O/H vs. $R_{\mathrm{23}}$
relationship. Triangles pointing up show the gas-phase oxygen abundance
estimated in the high-metallicity regime with the $R_{\mathrm{23}}$
method. Triangles pointing down show the low-metallicity regime, the
two estimates associated to one galaxy are connected by a dotted line.
The metallicity assigned to each galaxy is plotted with a filled symbol.
The dashed line shows the relation from \citet{Lamareille:2004MNRAS.350..396L}.
The solid line is the bisector least-squares fit to our data.}

\label{ohmaglh}
\end{figure}

Indeed, after extensive testing and evaluation of other methods, we
reached the conclusion that the best way to do this is to use the
$L-Z$ relation itself. To do this we compare the two metallicities
given by the $R_{\mathrm{23}}$ method and take the \emph{closest}
one to the $L-Z$ relation derived for the 2dFGRS \citep{Lamareille:2004MNRAS.350..396L}
to be our metallicity estimate for the galaxy (see Fig.~\ref{ohmaglh}).
We emphasize that we do \emph{not} assign a metallicity based on the
luminosity, and as we will discuss further below, this approach for
breaking the metallicity degeneracy is robust to rather substantial
changes in the $L-Z$ relation with redshift. This way of breaking
the degeneracy in the O/H vs. $R_{\mathrm{23}}$ relationship is called
the $L$ diagnostic in the rest of the paper.

Taking into account the high dispersion of the $L-Z$ relation (i.e.
rms = $0.25$ dex), we have to be very careful when using this diagnostic
in the turnaround region (i.e. $12+\log(\textrm{O/H})\approx8.4$)
of the O/H vs. $R_{\mathrm{23}}$ relationship, where the abundances
derived in the two regimes are both close enough to the $L-Z$ relation
to be kept. These have to be considered uncertain, but the fact that
the low- and high-metallicity points are equally distant from the
standard $L-Z$ relation indicates that the choice of metallicity
value \emph{will not significantly change the results} on the derived
$L-Z$ relation, in the intermediate metallicity regime. We have verified
that the $L-Z$ relation obtained when using only galaxies with a
N2 diagnostic or a reliable $L$ diagnostic is the same as the one
derived with all the points (difference $<1$\%).

\begin{figure}
\begin{center}\includegraphics[%
  width=1.0\columnwidth,
  keepaspectratio]{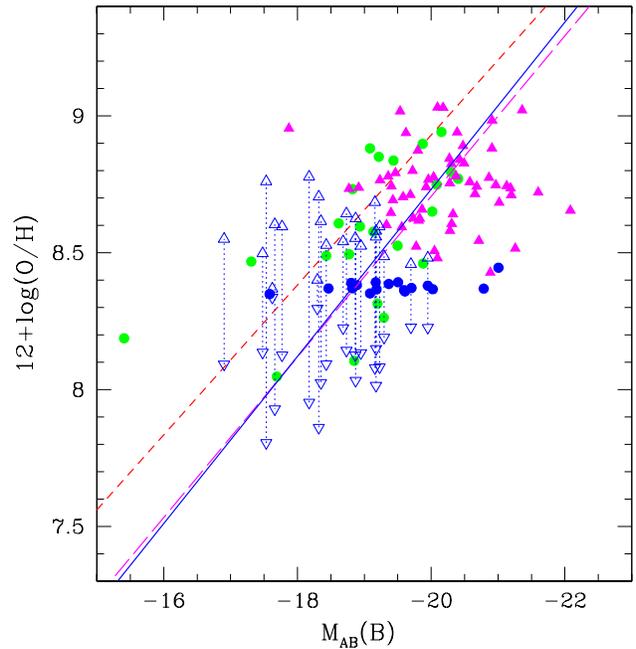}\end{center}

\caption{The Luminosity -- Metallicity relation with unreliable metallicity
estimates using the $R_{\mathrm{23}}$ method. Filled circles are
galaxies with an available N2 diagnostic for the metallicity determination.
Solid triangles are objects with a reliable $L$ diagnostic, and open
triangles are objects with an unreliable diagnostic (for these objects
the two possible abundances are plotted). The solid line shows the
linear regression on all these points, compared to the relation in
the local universe (short-dashed line) and the relation derived with
a reliable metallicity diagnostic (long-dashed line).}

\label{ohunsure}
\end{figure}

To illustrate this method we show in Fig.~\ref{ohunsure} the effect
of the $L$-diagnostic on the $L-Z$ relation. We present for each
galaxy the two estimates of $12+\log(\textrm{O/H})$ with the $R_{\mathrm{23}}$
method --- the upper branch is shown by upwards pointing triangles
and the lower branch by downwards pointing triangles. These are connected
by a dotted line. The metallicity assigned to a given galaxy by the
$L$-diagnostic is indicated with a filled symbol. The resulting $L-Z$
relation is still very similar to the standard relation (the new slope
value is $-0.31$) but with a larger dispersion (rms of the residuals
$=0.31$ dex). In view of these results, \emph{we conclude that the
possible errors introduced by the $L$ diagnostic, have a marginal
effect on the determination of the $L-Z$ relation at intermediate
redshifts}.

To be complete, we must point out that the $L-Z$ relation derived
using local galaxies may not be valid for intermediate-redshift galaxies.
However, previous works \citep{Kobulnicky:2003ApJ...599.1006K,Lilly:2003ApJ...597..730L}
have shown that the $L-Z$ relation has the same or \emph{steeper}
slope at high redshifts. A steeper slope would \emph{increase} the
difference between the $L-Z$ relation and the unselected abundance
estimate, strengthening the $L$ diagnostic for metallicity determinations.
In summary, the $L-Z$ relation derived using the $L$ diagnostic
\emph{will not} be biased if: \emph{i)} the $L-Z$ relation for our
sample of intermediate-redshift galaxies is the same as the $L-Z$
relation in the local universe, or \emph{ii)} the $L-Z$ relation
for our sample is steeper than the local $L-Z$ relation, or \emph{iii)}
the $L-Z$ relation for our sample has the same slope as the local
one but is shifted towards lower metallicities by less than $0.5$
dex. The other cases (i.e. a flatter $L-Z$ relation and an important
shift towards lower metallicities) cannot be derived without using
a non-degenerate metallicity indicator.

\subsubsection{Results\label{sub:ohResults}}

The results of the gas-phase oxygen abundances, estimated with the
$R_{\mathrm{23}}$ method, are shown in Table~\ref{ohdata} both
for the lower and upper branches of the O/H vs. $R_{\mathrm{23}}$
relationship. For the convenience of the reader, the redshift and
the absolute magnitude in the $B$-band are also given in this table.
Starting with 131 star-forming galaxies (selection described in Paper
I), the metallicity has been estimated for 121 of them (the 10 remaining
spectra do not show the {[}O\noun{ii}{]}$\lambda$3727 emission
line or give incompatible results). The \emph{final} adopted gas-phase
oxygen abundance together with its error estimate are reported in
the last column of Table~\ref{ohdata}. These values are the results
of the two methods used to break the O/H degeneracy as described in
previous sections. The N2 diagnostic is first used on 24 spectra to
determine the metallicity regime, then the $L$ diagnostic is used
on the 80 remaining spectra. Among them, 23 still have an unreliable
metallicity diagnostic. We have flagged those determinations in Table~\ref{ohdata}.

For a few galaxies, located in the turnaround region of the O/H vs.
$R_{\mathrm{23}}$ relationship, the O/H estimate given by the lower-branch
is \emph{higher} than that given by the upper branch (see Fig.~\ref{ohcalib}).
This occurs when the measured $R_{\mathrm{23}}$ parameter reaches
a higher value than the maximum allowed by the photo-ionization model.
The cause of this could be observational uncertainty mainly due to
a weak H$\beta$ emission line (which causes the high value of $R_{\mathrm{23}}$
and the big error bars shown in Fig.~\ref{ohcalib}) or problems
with the adopted calibrations. In any case we can not resolve this
issue and have decided to throw out the galaxies for which the difference
between the low and high estimates is big; in other cases the final
gas-phase oxygen abundance was computed as an average of these two
estimates; they are called {}``intermediate-metallicity'' galaxies.
We note that a higher proportion of the candidate star-forming galaxies
falls into the intermediate-metallicity region, because their $R_{23}$
parameter reaches by definition a high value (see Paper~I for a detailed
discussion on the spectal classification of the candidate star-forming
galaxies).

Four galaxies of our sample have previous gas-phase oxygen abundance
estimates from the litterature \citep{Lilly:2003ApJ...597..730L,Liang:2004A&A...423..867L,Maier:2005astro.ph..8239M},
as shown in Table~\ref{tabcompcfrs}. The values are all in relatively
good agreement. Unfortunatelly we do not have enough objects in common
to conclude in any bias between the different methods.

In summary, our sample of star-forming galaxies contains 10 low-metallicity
objects (8.3\%), 94 high-metallicity objects (77.7\%) and 17 intermediate-metallicity
objects (14.0\%).

\begin{table*}

\caption{Comparison between the gas-phase oxygen abundances found by us ($R_{\mathrm{23}}$
and CL01 methods) and in the litterature \citep{Lilly:2003ApJ...597..730L,Liang:2004A&A...423..867L,Maier:2005astro.ph..8239M}
for $4$ CFRS galaxies.}

\scriptsize{\label{tabcompcfrs}

\begin{tabular}{lllllll}
\hline\hline
CFRS	&LCL05	&$R_\mathrm{23}$	&CL01	&Lilly	&Liang	&Maier	\\
\hline
\object{22.0919}	&\object{045}	&8.48$\pm$0.01	&8.40$\pm$0.17	&8.30$\pm$0.20	&...	&8.38$\pm$0.14	\\
\object{03.0085}	&\object{030}	&8.82$\pm$0.16	&8.74$\pm$0.18	&8.84$\pm$0.07	&...	&8.36$\pm$0.41	\\
\object{03.0507}	&\object{023}	&8.77$\pm$0.03	&8.80$\pm$0.12	&...	&8.55$\pm$0.05	&...	\\
\object{03.0488}	&\object{022}	&8.71$\pm$0.06	&8.65$\pm$0.13	&...	&...	&8.88$\pm$0.07	\\
\hline
\end{tabular}
}
\end{table*}

\subsection{The \citet{Charlot:2001MNRAS.323..887C} method}

The standard $R_{\mathrm{23}}$ method, used in numerous studies to
estimate the oxygen abundance of nearby galaxies, is based on empirical
calibrations derived using photo-ionization models (see e.g. \citealp{McGaugh:1991ApJ...380..140M}).
An alternative would be to directly compare the observed spectra with
a library of photo-ionization models. The main advantage of such a
method is the use of more emission lines than the $R_{\mathrm{23}}$
method in a consistent way, which can give a better determination
of the metallicity. Here we will use a grid based on the models of
CL01. The CL01 models combine population synthesis models from \citet[version BC02]{Bruzual:2003MNRAS.344.1000B}
with emission line modelling from Cloudy \citep{Ferland:2001PASP..113...41F}
and a dust prescription from \citet{Charlot:2000ApJ...539..718C}.
The details of this grid are given in \citet{Brinchmann:2004MNRAS.351.1151B}
and Charlot et al. (in preparation), but we will summarize the most
important ones here for the convenience of the reader.

The models are parametrized by the total metallicity, $\log(Z)$,
the ionization parameter, $\log(U)$, the dust-to-metal ratio, $\xi$,
and the total dust attenuation, $\tau_{V}$. For each parameter the
model predicts the flux of an emission line and we compare these predictions
to our observed fluxes using a standard $\chi^{2}$ statistic. The
emission lines used here are {[}O\noun{ii}{]}$\lambda$3727, {[}O\noun{iii}{]}$\lambda\lambda$4959,5007,
{[}N\noun{ii}{]}$\lambda$6584, {[}S\noun{ii}{]}$\lambda\lambda$6717,6731,
H$\alpha$ and H$\beta$. The result of this procedure is a likelihood
distribution $P(\boldmath{Z,U,\xi,\tau_{V}})$ which can be projected
onto $12+\log(\textrm{O/H)}$ to construct the marginalized likelihood
distribution of $12+\log(\textrm{O/H})$.

\begin{figure}
\begin{center}\includegraphics[%
  width=1.0\columnwidth]{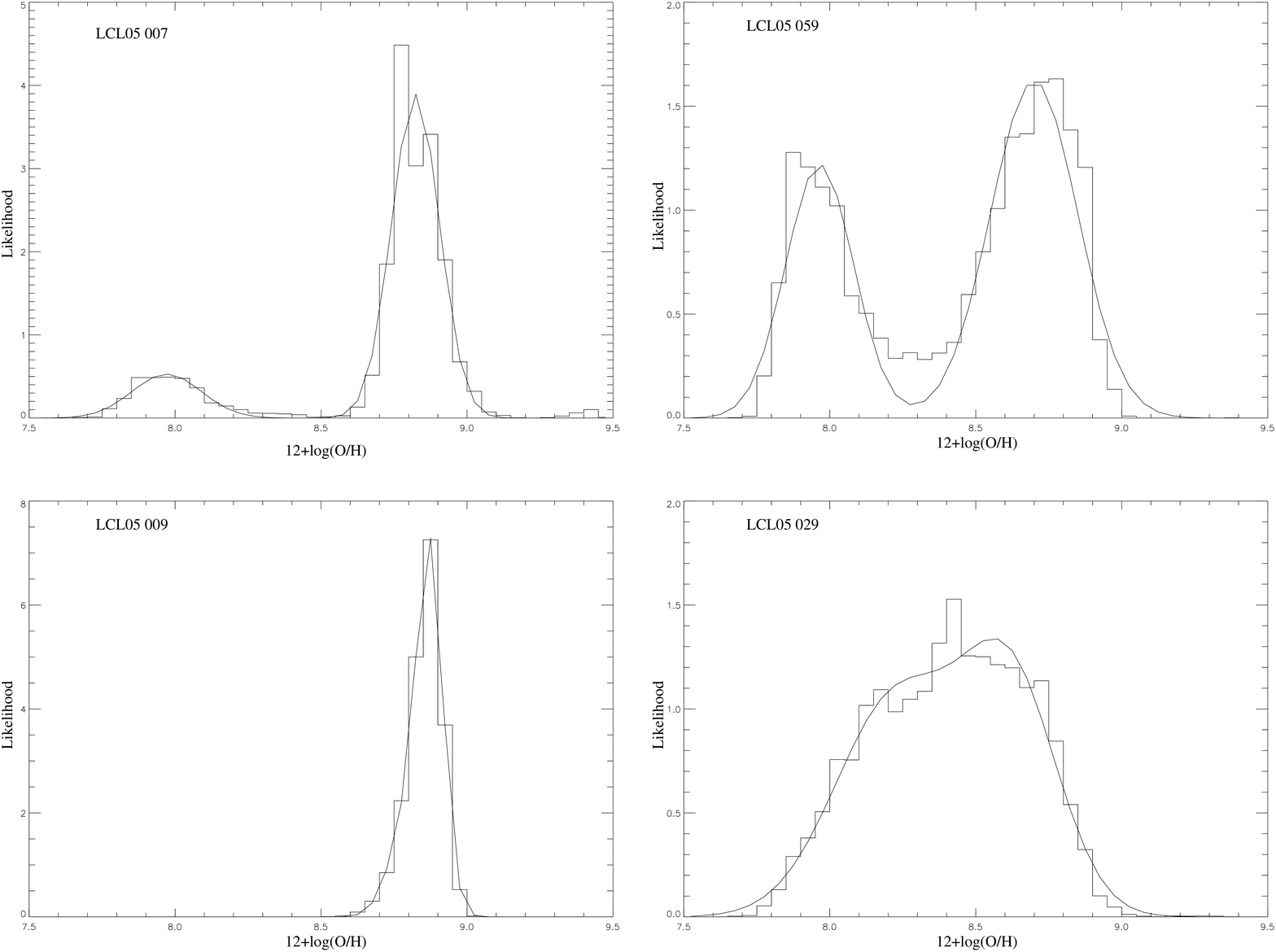}\end{center}

\caption{Examples of likelihood distributions of the gas-phase oxygen abundance,
and the associated gaussian fits, obtained with the \citet{Charlot:2001MNRAS.323..887C}
method. \textit{Top-left panel:} a degenerate case highly likely to
be a metal-rich galaxy (\object{LCL05\,007}, all lines used, faint
{[}O\noun{iii}{]}$\lambda$5007 line). \textit{Top-right panel:}
another degenerate case but with a much higher uncertainty on the
metallicity regime (\object{LCL05\,059}, blue lines only). \textit{Bottom-left
panel:} a non-degenerate case with a high metallicity (\object{LCL05\,009},
all lines used). \textit{Bottom-right panel:} a galaxy with an intermediate
metallicity (\object{LCL05\,029}, blue lines only, faint {[}O\noun{iii}{]}$\lambda$5007
line).}

\label{ohfitscl}
\end{figure}

We show some examples of these likelihood distributions in Fig.~\ref{ohfitscl}.
For the majority of our spectra we find double-peaked likelihood distributions
which is just a reflection of the inherent degeneracy of the strong-line
models when insufficient information is present. This is discussed
in more detail by Charlot et al (in preparation) who show that the
degeneracy is lifted with the inclusion of {[}N\noun{ii}{]}$\lambda$6584,
as for the standard $R_{\mathrm{23}}$ method as discussed above.

The Bayesian approach used in our model fits offers two alternative
methods to break the degeneracy in the O/H determinations. The first
is to use the global maximum of the likelihood distribution. The complex
shape of the likelihood surface makes this method rather unreliable
so we will not pursue this further, although for the bulk of the galaxies
it gives results consistent with the other methods. The second method
is to use the maximum-likelihood estimate which we derive by fitting
gaussians to each of the peaks of the likelihood distributions. To
do this we assume that the double-peaked likelihood distribution is
a combination of two nearly Gaussian distributions which correspond
to the two possible metallicities. We fit Gaussians to each peak and
take the mean of the Gaussian with the highest integrated probability
as our estimate of $12+\log(\textrm{O/H})$, with an error estimate
given by the sigma spread of the Gaussian fit.

\begin{figure}
\begin{center}\includegraphics[%
  width=1.0\columnwidth]{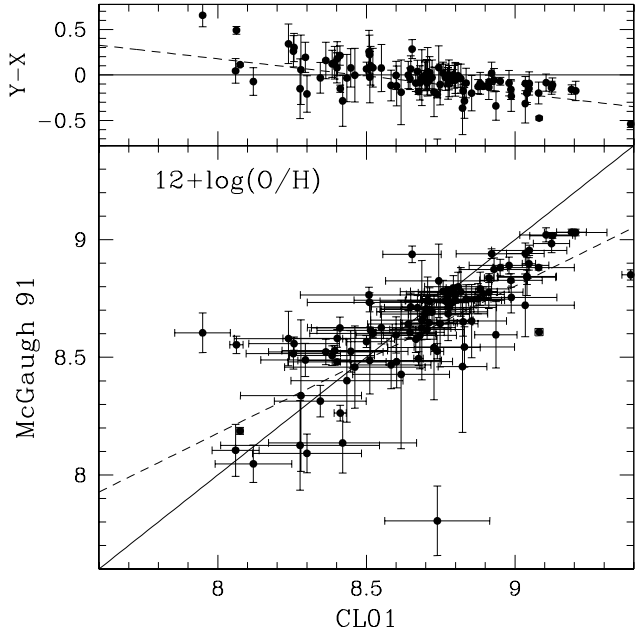}\end{center}

\caption{Comparison between the $R_{\mathrm{23}}$ and the CL01 estimates
of the metallicity. The top panel shows the residuals around the $y=x$
curve (solid line). The dashed line shows the linear fit.}

\label{compcl}
\end{figure}

Results of gas-phase oxygen abundance estimates with the CL01 method
are shown in Table~\ref{ohdatacl}. In Fig.~\ref{compcl}, we compare
the O/H estimates obtained by this method with those found by the
$R_{\mathrm{23}}$ method. The two values agree quite well, though
the methods adopted to break the O/H degeneracy and in how they deal
with the effect of dust attenuation (the CL01 method takes dust attenuation
into account in a self-consistent way, whereas the $R_{\mathrm{23}}$
method avoids the problem of dust attenuation by using EWs) are significantly
different. It is clear that there is a residual systematic difference
between the two estimators. A linear fit of the residuals shows a
slope of $0.62$. Only one object (\object{LCL05\,081}) show very
different results because of a contradictory metallicity classification
between the two methods (i.e. a low metallicity with the $L$ diagnostic,
but a high metallicity with the CL01 method). We acknowledge that
two different calibrations have different systematic uncertainties.
In this particular case, the trend to find higher metallicities with
the CL01 method is likely explained by the depletion of heavy elements
into dust grains, which is not taken into account by the \citet{McGaugh:1991ApJ...380..140M}
models.

\section{The Luminosity -- Metallicity relation\label{sec:lz}}

In this section, we first derive the luminosity -- metallicity relation
for our sample of intermediate-redshift galaxies using the gas-phase
oxygen abundances estimated with the $R_{\mathrm{23}}$ method and
the absolute magnitudes in the $B$ band, both listed in Table~\ref{ohdata}.
The linear regression used to derive the $L-Z$ relation is based
on the OLS linear bisector method \citep{Isobe:1990ApJ...364..104I},
it gives the bisector of the two least-squares regressions x-on-y
and y-on-x. See Appendix~\ref{sec:fitmethod} (only available online)
for a detailed discussion on the dependence of the $L-Z$ relation
on the fitting method.

\begin{figure}
\begin{center}\includegraphics[%
  width=1.0\columnwidth,
  keepaspectratio]{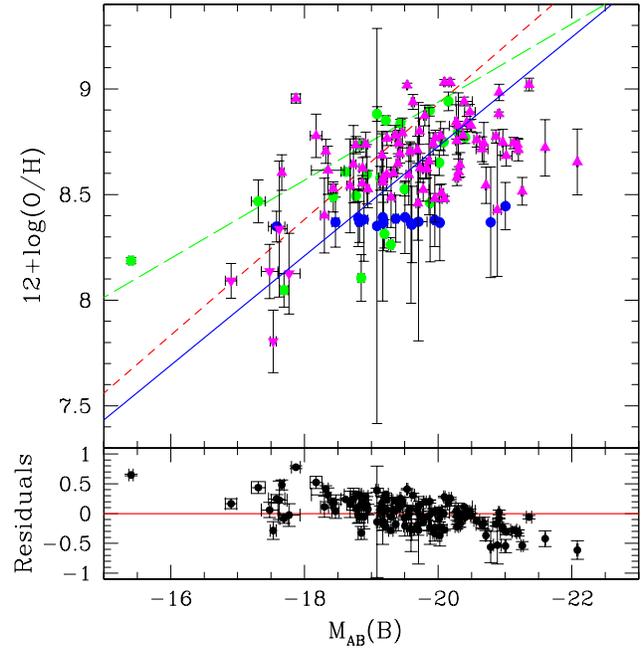}\end{center}

\caption{The Luminosity -- Metallicity relation in the $B$ band with gas-phase
oxygen abundances derived with the $R_{\mathrm{23}}$ method. The
filled circles are galaxies for which the metallicity regime was determined
using the N2 diagnostic, blue circles are objects in the intermediate
metallicity regime. In the other cases, the upwards pointing triangles
indicate objects that were determined to lie on the upper-branch and
the downwards pointing triangles those that lie on the lower branch.
The solid line shows the linear regression on the intermediate-redshift
galaxy sample. The short-dashed and long-dashed lines are the $L-Z$
relations in the local universe derived by \citet{Lamareille:2004MNRAS.350..396L}
(2dFGRS data) and \citet{Tremonti:2004astro.ph..5537T} (SDSS data)
respectively. The bottom panel shows the residuals around the solid
line.}

\label{ohb}
\end{figure}

Fig.~\ref{ohb} shows the $L-Z$ relation based on the N2 diagnostic
to break the O/H degeneracy, whenever possible, and on the $L$ diagnostic
otherwise. We obtain the following relation in the $B$ band: \begin{equation}
12+\log(\textrm{O/H})=-(0.26\pm0.03)M_{AB}(B)+3.55\pm0.54\label{eq:lz1}\end{equation}
 with a rms of the residuals of $0.25$ dex. This dispersion is very
close to the one obtained for the $L-Z$ relation in the local universe
using 2dFGRS data and derived by \citet{Lamareille:2004MNRAS.350..396L}.
The existence of the $L-Z$ relation at intermediate redshifts is
clearly confirmed.

We also derive a $L-Z$ relation in the $R$ band, using magnitudes
reported in Paper I: \begin{equation}
12+\log(\textrm{O/H})=-(0.24\pm0.03)M_{AB}(R)+3.82\pm0.52\label{eq:lz2}\end{equation}
 with a rms of the residuals of $0.26$ dex, still very similar to
the $R$-band $L-Z$ relation in the local universe \citep{Lamareille:2004MNRAS.350..396L}.
Note however that the dispersion is not smaller than for the relation
in the $B$ band. We will thus focus our analysis on the $B$-band
$L-Z$ relation in order to allow comparisons with previous studies
for which the $B$-band magnitudes are commonly used.

\begin{figure}
\begin{center}\includegraphics[%
  width=1.0\columnwidth]{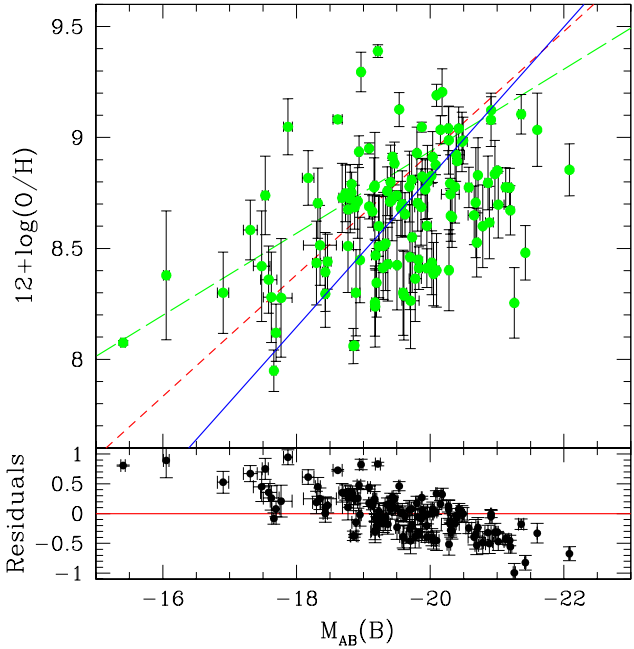}\end{center}

\caption{The $L-Z$ relation with the CL01 estimate of the metallicity. The
short-dashed line is the $L-Z$ relation in the local universe derived
with the 2dFGRS data \citep{Lamareille:2004MNRAS.350..396L}. The
long-dashed line is the same relation derived with the SDSS data \citep{Tremonti:2004astro.ph..5537T}.
The solid line is the $L-Z$ relation derived for our sample of intermediate-redshift
galaxies. The bottom panel shows the residuals around the solid line.}

\label{lzcl}
\end{figure}

We then derive the luminosity -- metallicity relation for intermediate-redshift
galaxies using the gas-phase oxygen abundances estimated with the
CL01 method (reported in Table~\ref{ohdatacl}) and the $B$-band
absolute magnitudes listed in Table~\ref{ohdata}. We find the following
relation (see Fig.~\ref{lzcl}): \begin{equation}
12+\log(\textrm{O/H})=-(0.34\pm0.05)M_{AB}(B)+2.06\pm0.99\label{eq:lzcl1}\end{equation}
 with a rms of the residuals of $0.36$ dex.

We conclude that the $L-Z$ relation at intermediate redshifts, obtained
with the CL01 estimate of the metallicity, is not significantly different
from the local $L-Z$ relation, considering that we break the degeneracy
between low- and high-metallicity estimates with the maximum-likelihood
method, which adds some uncertainties.

\subsection{Selection effects}

In this section we investigate the different selection effects which
could introduce systematic biases in the $L-Z$ relation. First of
all, the sample selection, based on the presence of emission lines
in the galaxy spectra (see Paper~I for details), introduces a straight
cut at high metallicities, where the oxygen lines become too faint
to be measured. The consequence of this effect is that all galaxies
have a gas-phase oxygen abundance $12+\log(\textrm{O/H})<9.0$. Another
bias introduced by the redshift is the result of the selection applied
for the CFRS sub-sample: galaxies in the redshift range $0.2<z<0.4$
were preferred in order to observe the {[}N\noun{ii}{]}$\lambda$6584
and H$\alpha$ emission lines.

\begin{figure}
\begin{center}\includegraphics[%
  width=1.0\columnwidth]{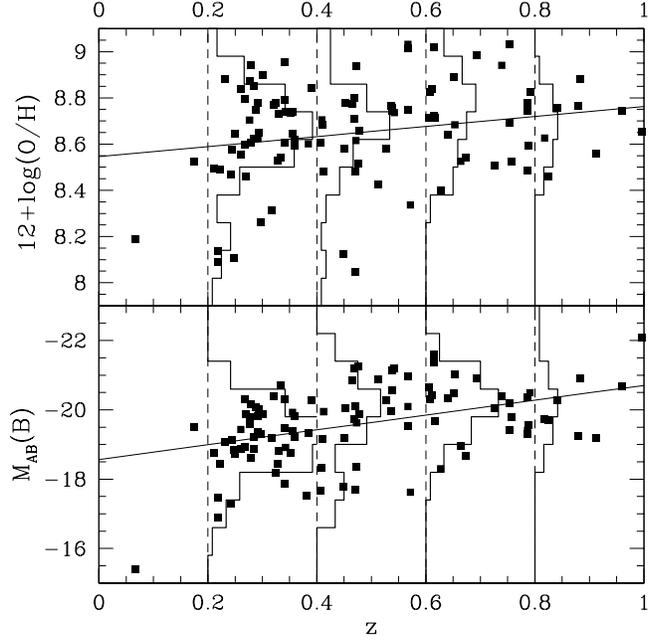}\end{center}

\caption{A view of our sample's selection effects. This figure shows, as a
function of redshift, the distribution of the absolute $B$-band magnitudes
(bottom panel) and metallicities (top panel) of our sample of intermediate-redshift
galaxies. The straight line shows a linear regression of the observed
(not real) evolution. Histograms are plotted for four redshift bins:
$0.2\leq z\leq0.4$, $0.4<z\leq0.6$, $0.6<z\leq0.8$ and $z>0.8$.}

\label{effselect}
\end{figure}

Since our sample is limited in apparent magnitude, the Malmquist bias,
whereby the minimum luminosity observed increases with redshift, is
clearly present (see Fig.~\ref{effselect}). This turns out to be
the most problematic bias. This also affects the metallicity which
is linked to the luminosity through the $L-Z$ relation. This bias
results in an apparent increase of the observed metallicity with increasing
redshift whereas galaxy evolution models predict that, on average,
the metallicity should decrease with increasing redshift. Nevertheless,
Fig.~\ref{effselect} shows that our sample of intermediate-redshift
galaxies seems to be complete in the redshift range $0.2<z<0.6$,
allowing us to perform reliable comparisons with local samples.

We checked that there is no significant bias between the three sub-samples
(CFRS, CLUST and GDDS sub-samples) considered in this study.

\subsection{Addition of other samples of intermediate-redshift galaxies\label{sub:Addition-of-other}}

In order to study the $L-Z$ relation for a larger and more complete
(in terms of redshift coverage) sample of intermediate-redshift galaxies,
we have performed an exhaustive compilation of star-forming galaxies
with relevant data (luminosity and metallicity derived with the $R_{\mathrm{23}}$
method) available in the literature \citep{Kobulnicky:1999ApJ...511..118K,Hammer:2001ApJ...550..570H,Contini:2002MNRAS.330...75C,Kobulnicky:2003ApJ...599.1006K,Lilly:2003ApJ...597..730L,Liang:2004A&A...423..867L}.
The luminosities were adjusted to our adopted cosmology and converted
to the AB system when necessary.

\begin{figure}
\begin{center}\includegraphics[%
  width=1.0\columnwidth]{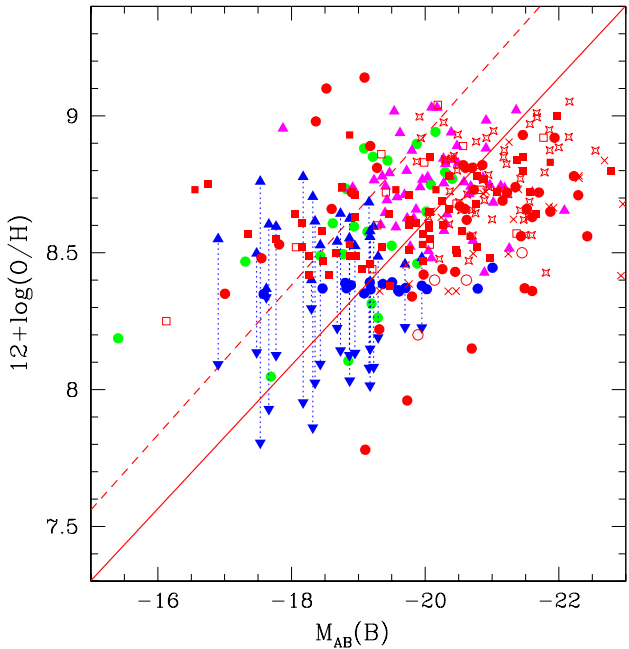}\end{center}

\caption{The $L-Z$ relation including additional samples of intermediate-redshift
galaxies and using metallicities derived with the $R_{\mathrm{23}}$
method. Green filled circles are objects where N2 diagnostic is used
for metallicity estimate, magenta triangles are objects with a secure
$L$ diagnostic, and blue triangles are objects with an unreliable
diagnostic (for these objects the two possible abundances are shown).
Other samples of intermediate-redshift galaxies drawn from the literature
are plotted in red: open squares from \citet{Kobulnicky:1999ApJ...511..118K},
solid squares from \citet{Kobulnicky:2003ApJ...599.1006K}, open circles
from \citet{Hammer:2001ApJ...550..570H}, filled circles from \citet{Liang:2004A&A...423..867L},
stars from \citet{Lilly:2003ApJ...597..730L} and crosses from \citet{Contini:2002MNRAS.330...75C}
($z>0.1$). The solid line shows the linear regression for all these
points, compared to the relation obtained in the local universe from
2dFGRS data (dashed line).}

\label{ohmagcompz}
\end{figure}

The $L-Z$ relation for this extended sample, and using metallicities
derived with the $R_{\mathrm{23}}$ method, takes the following form:
\begin{equation}
12+\log(\textrm{O/H})=-(0.26\pm0.03)M_{AB}(B)+3.36\pm0.53\label{eq:lz3}\end{equation}
 with a rms of the residuals of $0.33$ (see Fig.~\ref{ohmagcompz}).

\begin{figure}
\begin{center}\includegraphics[%
  width=1.0\columnwidth]{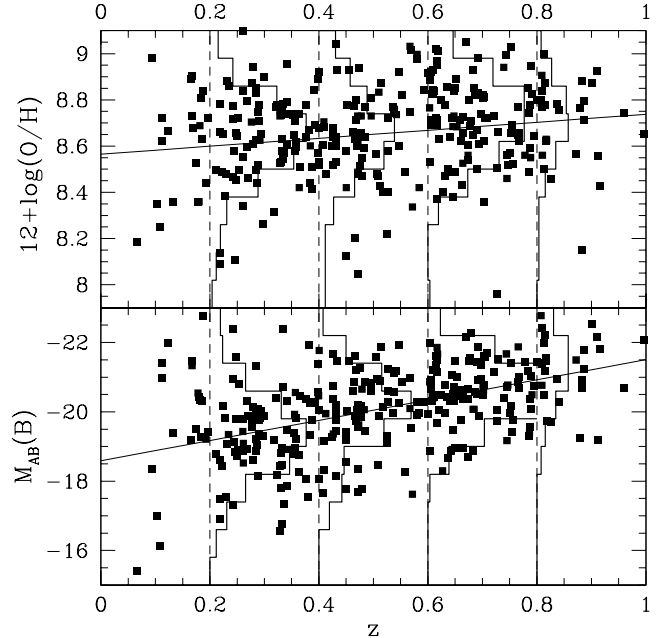}\end{center}

\caption{Selection effects when we include other samples of intermediate-redshift
galaxies drawn from the literature (see text for details). This figure
shows, as a function of redshift, the distribution of the absolute
$B$-band magnitudes (bottom panel) and metallicities (top panel)
of the combined sample of intermediate-redshift galaxies. The straight
line shows the linear regression of the observed (not real) evolution.
Histograms are plotted for four redshift bins: $0.2\leq z\leq0.4$,
$0.4<z\leq0.6$, $0.6<z\leq0.8$ and $z>0.8$.}

\label{effselectcomp}
\end{figure}

The $L-Z$ relation at intermediate redshifts is again very close
to the one in the local universe derived from 2dFGRS data. It has
almost the same slope and zero-point than for the local relation,
while the metallicity is on average $0.32$ dex lower. This difference
is within the uncertainty associated to the zero-point ($0.53$ dex).
However, this difference in the zero point has already been interpreted
in previous studies as a deficiency by a factor of 2 for the metallicity
of intermediate-redshift galaxies compared to local samples \citep{Hammer:2004astroph0410518,Kobulnicky:2004astro.ph.10684K}.
We also note that this extended sample is clearly biased towards high-luminosity
objects at high redshifts, as shown by the histograms in Fig.~\ref{effselectcomp}.

We conclude that the $L-Z$ relation, derived for the \emph{whole}
sample of intermediate-redshift galaxies (i.e. our own sample combined
with other data drawn from the literature), is still very similar,
in term of slope, to the $L-Z$ relation in the local universe. The
possible shift in the zero point between the two $L-Z$ relations
will be investigated further in Sect.~\ref{sec:evolz}.

\subsection{Different regimes of the $L-Z$ relation\label{sec:residuals}}

Different authors (e.g. \citealp{Melbourne:2002AJ....123.2302M,Lamareille:2004MNRAS.350..396L})
have suggested that the global $L-Z$ relation in the local universe
may not be simply approximated by a \emph{single} linear relation
and proposed to divide the galaxy samples into metal-poor dwarfs and
metal-rich spirals. The $L-Z$ relations derived for these two subsamples
by \citet{Lamareille:2004MNRAS.350..396L} are indeed different, with
metal-rich galaxies following a steeper $L-Z$ relation than the metal-poor
ones.

We can try to apply this distinction between metal-poor and metal-rich
galaxies to our samples of intermediate-redshift galaxies, in order
to derive {}``metallicity-dependant'' $L-Z$ relations. However,
we believe that a constant limit in metallicity to separate the two
galaxy populations (e.g. $12+\log(\textrm{O}/\textrm{H})=8.3$ as
used in \citealp{Lamareille:2004MNRAS.350..396L}) is not the best
way to do as it does not take into account accurately the high dispersion
of observed metallicities at a given galaxy luminosity. Indeed there
is no clear physical separation between galaxies having a metallicity
lower or higher than one specific value.

Instead, taking into account the intrinsic scatter of the $L-Z$ relation,
we try in this section to define a {}``region fitting'' which gives
the behavior of the $L-Z$ relation in a metal-rich and a metal-poor
regimes (i.e. respectively the upper and the lower regions of the
$L-Z$ relation). We assume that the separation between metal-rich
and metal-poor galaxies increases with the absolute magnitude with
a similar slope of $-0.27$ than for the $L-Z$ relation in the local
universe from the 2dFGRS. \emph{The separation between metal-poor
and metal-rich galaxies is thus equal to the mean metallicity of a
galaxy at a given luminosity}. Note that this method is not sensitive
to whether the physical mixing of galaxies in the $L-Z$ diagram is
done along the x-axis or along the y-axis (if the mixing is done along
the x-axis, this is rather a high/low-luminosity separation).

We emphasize that we do not use the standard, horizontal, separation
between low- and high-metallicity objects described in Sect.~\ref{sub:Breaking-the-degeneracy}.
We use instead a luminosity-dependant separation which gives us $61$
metal-poor and $59$ metal-rich galaxies. Please note that in order
to take into account the lower average metallicity of our sample (see
Fig.~\ref{ohb}), we have shifted our luminosity-dependent separation
between metal-rich and metal-poor galaxies towards lower metallicities
by $-0.20$ dex, keeping the same slope given by the local $L-Z$
relation.

\begin{figure}
\begin{center}\includegraphics[%
  width=1.0\columnwidth]{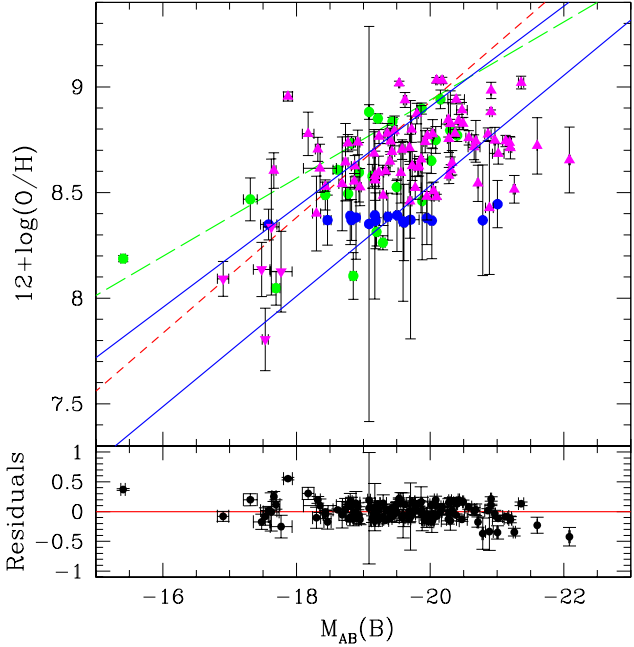}\end{center}

\caption{The {}``region-fitted'' $L-Z$ relations for our own sample of
intermediate-redshift galaxies. The metallicities are derived with
the $R_{\mathrm{23}}$ method. The solid lines represent the two linear
regressions for the metal-poor and metal-rich galaxies (see text for
details). Same legend as in Fig.~\ref{ohb} for the data points.
The bottom panel shows the residuals around these two fits.}

\label{lzcl2}
\end{figure}

We have applied this method to divide our sample of intermediate-redshift
galaxies into metal-rich and metal-poor galaxies using the $L-Z$
relation derived in the local universe from the 2dFGRS data. The two
new linear fits obtained for each sub-sample are shown in Fig.~\ref{lzcl2}.
We derive the following relations:

\begin{equation}
12+\log(\textrm{O/H})=-(0.26\pm0.02)M_{AB}(B)+3.30\pm0.38\label{eq:lzcl2a}\end{equation}
 for metal-poor galaxies, and

\begin{equation}
12+\log(\textrm{O/H})=-(0.24\pm0.03)M_{AB}(B)+4.15\pm0.49\label{eq:lzcl2b}\end{equation}
 for metal-rich galaxies. The rms of the residuals for the whole sample
is now equal to $0.15$ dex only, and we see on the bottom panel of
Fig.~\ref{lzcl2} that there is almost no residual slope. This has
to be compared with the quite high dispersion (rms $\sim0.28-0.34$
dex) and the non-zero slope of the residuals obtained for the \emph{single}
$L-Z$ relation (see Fig.~\ref{ohb}).

The same distinction between metal-poor and metal-rich galaxies can
be applied to the whole sample of intermediate-redshift galaxies,
i.e. adding the samples drawn from the literature (see Sect.~\ref{sub:Addition-of-other}).

\begin{figure}
\begin{center}\includegraphics[%
  width=1.0\columnwidth]{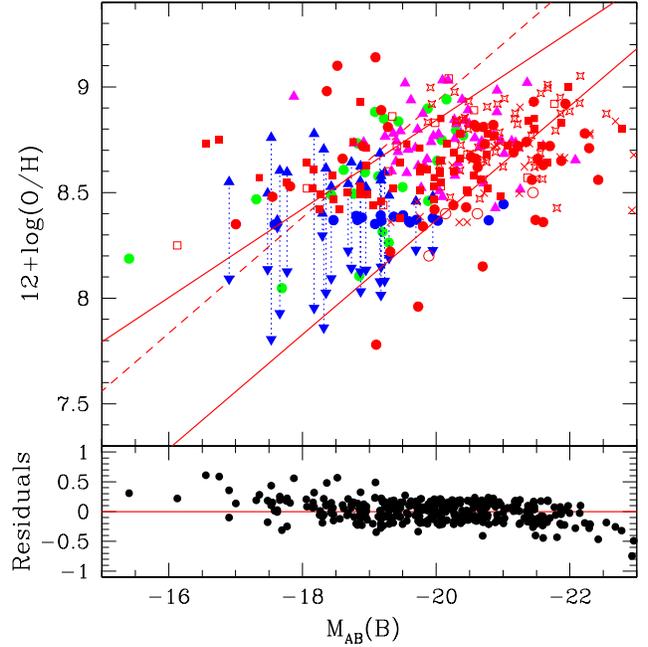}\end{center}

\caption{The {}``region-fitted'' $L-Z$ relations including additional samples
of intermediate-redshift galaxies (as described in Sect.~\ref{sub:Addition-of-other}).
The metallicities are derived with the $R_{\mathrm{23}}$ method.
The solid lines represent the two linear regressions for the metal-poor
and metal-rich galaxies (see text for details). Same legend as in
Fig.~\ref{ohb} for the data points. The bottom panel shows the residuals
around these two fits.}

\label{lzcompz2}
\end{figure}

The two new linear fits obtained for each sub-sample (the separation
is shifted by $-0.32$ dex) are shown in Fig.~\ref{lzcompz2}. We
derive the following relations:

\begin{equation}
12+\log(\textrm{O/H})=-(0.27\pm0.03)M_{AB}(B)+2.96\pm0.53\label{eq:lzcl2a}\end{equation}
 for $184$ metal-poor galaxies , and

\begin{equation}
12+\log(\textrm{O/H})=-(0.21\pm0.01)M_{AB}(B)+4.65\pm0.29\label{eq:lzcl2b}\end{equation}
 for $164$ metal-rich galaxies. The rms of the residuals for the
whole sample is equal to $0.19$ dex and we see again on the bottom
panel of Fig.~\ref{lzcompz2} that there is almost no residual slope.

The metallicity-dependent $L-Z$ relations derived above are very
similar, in terms of slope and zero point, if we consider our sample
alone or the whole sample of intermediate-redshift galaxies.

\subsection{Evolution of the $L-Z$ relation with redshift?\label{sec:evolz}}

We now investigate a possible evolution with redshift of the $L-Z$
relation, using the $R_{\mathrm{23}}$ method for the metallicity
determination.

We first divided our own sample of intermediate-redshift galaxies
into four redshift bins: $0.2\leq z\leq0.4$, $0.4<z\leq0.6$, $0.6<z\leq0.8$
and $0.8<z<1.0$.

\begin{figure}
\begin{center}\includegraphics[%
  width=1.0\columnwidth]{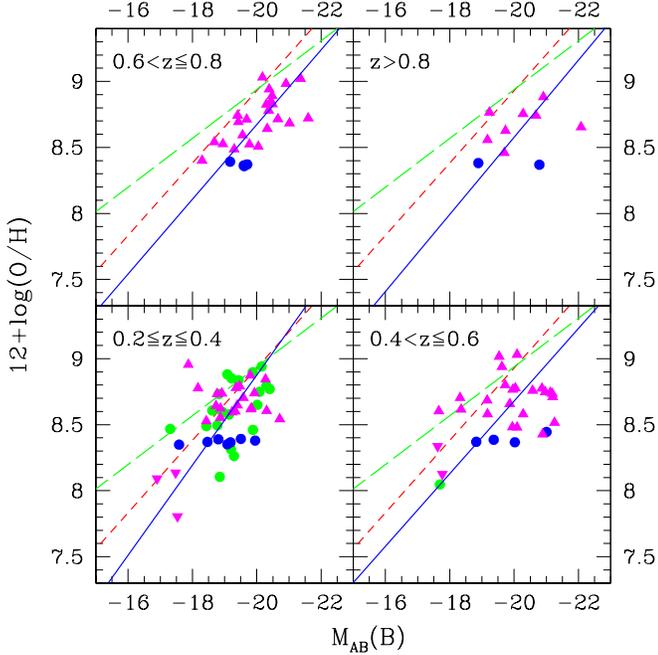}\end{center}

\caption{Redshift evolution of the $L-Z$ relation (see text for details).
The four panels show the $L-Z$ relation for our own sample of intermediate-redshift
galaxies in the redshift ranges $0.2\leq z\leq0.4$, $0.4<z\leq0.6$,
$0.6<z\leq0.8$ and $0.8<z<1.0$. The legend is the same as in Fig.~\ref{ohb}.}

\label{figevol1}
\end{figure}

The resulting $L-Z$ relations per redshift bin are shown in Fig.~\ref{figevol1}.
The parameters (slope, zero points) of the linear regressions are
listed in Table~\ref{tabevol1}. The parameter $f_{\mathrm{above}}$
corresponds to the fraction of galaxies located above the local $L-Z$
relation derived with 2dFGRS data.

With the present sample, we only see \emph{a small, not significant,
evolution of the $L-Z$ relation with redshift}, in terms of slope
and zero point, from the local universe to $z\sim1$. We only find
statistical variations consistent with the uncertainty in the derived
parameters. The small variation is however confirmed by a decreasing
fraction of galaxies falling above the local $L-Z$ relation ($f_{\mathrm{above}}$
in Table~\ref{tabevol1}). 

\begin{table}

\caption{Redshift evolution of the slope, zero point and rms of the residuals
of the $L-Z$ relation for our own sample of intermediate-redshift
star-forming galaxies. Metallicities have been estimated with the
$R_{\mathrm{23}}$ method. The parameter $f_{\mathrm{above}}$ represents
the fraction of galaxies located above the local $L-Z$ relation.
The results obtained on the 2dFGRS sample \citep{Lamareille:2004MNRAS.350..396L}
are given for reference.}

\label{tabevol1}

\begin{tabular}{lllll}
\hline 
\hline
Redshift bin&
Slope&
Zero Point&
rms&
$f_{\mathrm{above}}$\tabularnewline
\hline
2dFGRS&
$-0.27\pm0.01$&
$3.45\pm0.09$&
$0.27$&
\tabularnewline
$0.2\le z\le0.4$&
$-0.34\pm0.05$&
$2.06\pm0.98$&
$0.27$&
$0.45$\tabularnewline
$0.4<z\le0.6$&
$-0.27\pm0.06$&
$3.18\pm1.24$&
$0.30$&
$0.30$\tabularnewline
$0.6<z\le0.8$&
$-0.28\pm0.04$&
$3.01\pm0.85$&
$0.17$&
$0.15$\tabularnewline
$0.8<z<1.0$&
$-0.29\pm0.17$&
$2.74\pm3.44$&
$0.27$&
$0.10$\tabularnewline
\hline
\end{tabular}
\end{table}

\begin{figure}
\begin{center}\includegraphics[%
  width=1.0\columnwidth]{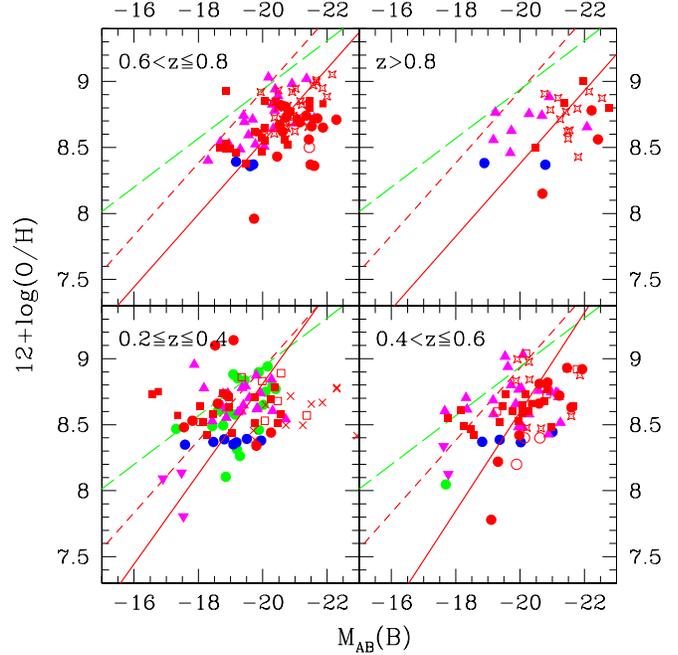}\end{center}

\caption{Redshift evolution of the $L-Z$ relation using the $R_{\mathrm{23}}$
method for the metallicity determination and including other samples
of intermediate-redshift galaxies (as described in Sect.~\ref{sub:Addition-of-other}).
The four panels show the $L-Z$ relation for the full combined sample
in the redshift ranges $0.2\leq z\leq0.4$, $0.4<z\leq0.6$, $0.6<z\leq0.8$
and $0.8<z<1.0$. The legend is the same as in Fig.~\ref{ohmagcompz}.}

\label{figevol2}
\end{figure}

In order to be more complete in searching for any possible evolution
of the $L-Z$ relation with redshift, Fig.~\ref{figevol2} and Table~\ref{tabevol2}
show the $L-Z$ relation with the addition of other samples of intermediate-redshift
galaxies (as described in Sect.~\ref{sub:Addition-of-other}) divided
into four redshift bins: $0.2\leq z\leq0.4$, $0.4<z\leq0.6$, $0.6<z\leq0.8$
and $0.8<z<1.0$.

As previously observed with our sample alone, the $L-Z$ relation
only shows a small evolution, in terms of slope and zero points, between
these four redshift bins. Again, we only find statistical variations
consistent with the uncertainty in the derived parameters (see Table~\ref{tabevol2}).

However, we remark that the $L-Z$ relation tends to shift toward
lower metallicities when the redshift increases, which is confirmed
by a decrease of the fraction of galaxies $f_{\mathrm{above}}$ above
the local $L-Z$ relation (from $43$\% at $z\sim0.3$ to $3$\% at
$z\sim1$, see Table~\ref{tabevol2}).

\begin{table}

\caption{Redshift evolution of the slope, zero point and rms of the residuals
of the $L-Z$ relation including other samples of intermediate-redshift
galaxies (as described in Sect.~\ref{sub:Addition-of-other}). Metallicities
have been estimated with the $R_{\mathrm{23}}$ method. The parameter
$f_{\mathrm{above}}$ represents the fraction of galaxies located
above the local $L-Z$ relation.}

\label{tabevol2}

\begin{tabular}{lllll}
\hline 
\hline
Redshift bin&
Slope&
Zero Point&
rms&
$f_{\mathrm{above}}$\tabularnewline
\hline
$0.2\le z\le0.4$&
$-0.34\pm0.10$&
$1.98\pm1.85$&
$0.43$&
$0.43$\tabularnewline
$0.4<z\le0.6$&
$-0.37\pm0.09$&
$1.10\pm1.86$&
$0.39$&
$0.22$\tabularnewline
$0.6<z\le0.8$&
$-0.28\pm0.04$&
$3.04\pm0.81$&
$0.23$&
$0.07$\tabularnewline
$0.8<z<1.0$&
$-0.28\pm0.08$&
$2.86\pm1.72$&
$0.27$&
$0.03$\tabularnewline
\hline
\end{tabular}
\end{table}

To better quantify this possible evolutionary effect, we estimate
for each galaxy with a given luminosity, the difference between its
metallicity and the one given by the $L-Z$ relation in the local
universe. For each redshift bin, the mean value of this difference
give us the \emph{average} metallicity shift, assuming that, in average,
the metallicity increases with the galaxy luminosity with a slope
equal to $-0.27$ (see Table~\ref{tabevol1}). The results are shown
in Table~\ref{tabevolcl} (case a). We clearly see a decrease of
the mean metallicity for a given luminosity when the redshift increases.
The last column shows that the galaxies in the redshift range $0.8<z<1.0$
appear to be metal-deficient by a factor of $\sim3$ compared with
galaxies in the local universe. For a given luminosity, they contain
on average about third of the metals locked in local galaxies.

We acknowledge that the evolution of the $L-Z$ relation is a combination
of a metallicity and a luminosity evolution at a given stellar mass.
Unfortunately our sample is not large enough to statistically distinguish
between the two effects, but for the convenience of the reader we
have done the same calculations after correcting the luminosity of
the galaxies from the evolution. The results are listed in Table~\ref{tabevolcl}
(case b). We have used the last results obtained on the evolution
of the galaxy luminosity function in the VVDS first epoch data \citep{Ilbert:2004astro.ph..9134I},
which give an average evolution in the $B$-band magnitude of $-0.3$
at $z=0.2$, $-0.7$ at $z=0.4$, $-0.9$ at $z=0.8$ and $-1.0$
at $z=1.0$ since $z=0.0$. If these values correctly reflect the
luminosity evolution of our sample, the metallicity decrease at redshift
up to $z=1.0$ would be of a factor $\sim1.3$ for a given stellar
mass (if we neglect the effect of the different mass-to-light ratios).

\begin{table}

\caption{Evolution with redshift of the mean value of the metallicity difference
with the $L-Z$ relation in the local universe. The average difference
of metallicity (derived with the $R_{\mathrm{23}}$ method) is computed
for four redshift bins for our sample of star-forming galaxies, including
other samples of intermediate-redshift galaxies (as described in Sect.~\ref{sub:Addition-of-other}).
Case (a) is uncorrected data, case (b) is after the luminosities have
been corrected from evolution (see text for details). The two last
columns show the same value converted into a linear scale. It indicates
the amount of metals in each redshift bin compared with the value
in the local universe.}

\label{tabevolcl}

\begin{tabular}{lllll}
\hline 
\hline
&
\multicolumn{2}{l}{$\left\langle Z-Z_{2dF}(L)\right\rangle $}&
\multicolumn{2}{l}{linear}\tabularnewline
Redshift bin&
(a)&
(b)&
(a)&
(b)\tabularnewline
\hline
$0.2\le z\le0.4$&
$-0.16$&
$-0.08$&
$0.97$&
$1.17$\tabularnewline
$0.4<z\le0.6$&
$-0.33$&
$-0.14$&
$0.60$&
$0.93$\tabularnewline
$0.6<z\le0.8$&
$-0.38$&
$-0.14$&
$0.47$&
$0.83$\tabularnewline
$0.8<z<1.0$&
$-0.55$&
$-0.28$&
$0.34$&
$0.64$\tabularnewline
\hline
\end{tabular}
\end{table}

\section{Conclusion}

Starting with a sample of 129 star-forming galaxies at intermediate
redshifts ($0.2<z<1.0$), for which the sample selection, the observations
and associated data reduction, the photometric properties, the emission-line
measurements, and the spectral classification are described in Paper~I,
we derived the gas-phase oxygen abundance O/H which is used as a tracer
of the metallicity. We used two methods: the $R_{\mathrm{23}}$ method
\citep{McGaugh:1991ApJ...380..140M} which is based on empirical calibrations,
and the CL01 method \citep{Charlot:2001MNRAS.323..887C} which is
based on grids of photo-ionization models and on fitting emission
lines. We have investigated the problem of the metallicity degeneracy
between the high- and low-metallicity branches of the O/H vs. $R_{\mathrm{23}}$
relationship. The following conclusions have been drawn from this
study:

\begin{itemize}
\item The N2 diagnostic based on the {[}N\noun{ii}{]}$\lambda6584$/H$\alpha$
line ratio is the best way to discriminate between high- and low-metallicity
objects. 
\item The $L$ diagnostic can be used with a relatively high confidence
level for galaxies far enough from the turnaround region (12+log(O/H)
$\sim8.3$) of the O/H vs. $R_{\mathrm{23}}$ relationship. Note however
that, in this region, the error in the final abundance will be low. 
\item Any diagnostic based on the galaxy color or the intensity of the Balmer
break will fail because of the high dispersion in the relations between
metallicity and these parameters. 
\item The CL01 method offers a good way to break the metallicity degeneracy
with the maximum likelihood method, even when the {[}N\noun{ii}{]}$\lambda6584$
and H$\alpha$ lines are not available. 
\end{itemize}
We have then derived the following luminosity -- metallicity ($L-Z$)
relations: the $L-Z$ relation in the $B$-band and in the $R$-band
using the $R_{\mathrm{23}}$ method for metallicity determinations
(equations~\ref{eq:lz1} and~\ref{eq:lz2}), the $L-Z$ relation
in the $B$-band with the CL01 method to derive metallicities (equation~\ref{eq:lzcl1}),
and the $L-Z$ relation for our galaxies combined with other samples
of intermediate-redshift galaxies drawn from the literature (see Sect.~\ref{sub:Addition-of-other})
and the $R_{\mathrm{23}}$ method for metallicity estimates (equation~\ref{eq:lz3}).
We investigated the possibility to divide, for a given luminosity,
the sample into metal-rich and metal-poor galaxies in order to do
a {}``region fitting'' instead of a single linear fit. We thus derived
two new $L-Z$ relations showing similar slopes but lower residuals
(equations~\ref{eq:lzcl2a} and~\ref{eq:lzcl2b}). We draw the following
conclusions from this analysis:

\begin{itemize}
\item The $L-Z$ relations at intermediate redshifts are very similar in
term of slope to the $L-Z$ relations obtained in the local universe
\citep{Lamareille:2004MNRAS.350..396L,Tremonti:2004astro.ph..5537T}. 
\item When including other samples of intermediate-redshift galaxies (see
Sect.~\ref{sub:Addition-of-other}), we find a $L-Z$ relation which
is shifted by $\sim0.3$ dex towards lower metallicities compared
with the local one.
\end{itemize}
Finally, we investigated any possible evolution of these $L-Z$ relations
with redshift. We find that:

\begin{itemize}
\item Our sample alone does not show any significant evolution of the $L-Z$
relation up to $z\sim1.0$. We only find statistical variations consistent
with the uncertainty in the derived parameters.
\item Including other samples of intermediate-redshift galaxies (see Sect.~\ref{sub:Addition-of-other}),
we clearly see, at a given galaxy luminosity, a decrease of the mean
metallicity when the redshift increases. Galaxies at $z\sim1$ appear
to be metal-deficient by a factor of $\sim3$ compared with galaxies
in the local universe. For a given luminosity, they contain on average
about one third of the metals locked in local galaxies.
\item If we apply a correction for the luminosity evolution, galaxies at
$z\sim1$ appear to be metal-deficient by a factor of $\sim1.6$ compared
with galaxies in the local universe. 
\end{itemize}
Further analysis of our sample of intermediate-redshift galaxies,
in terms of mass and star formation history, will be performed in
subsequent papers.

\begin{acknowledgements}
F.L. would like to thank warmly H. Carfantan for help and valuable
discussions about fitting methods, and E. Davoust for English improvement.
J.B. acknowledges the receipt of an ESA external post-doctoral fellowship.
J.B. acknowledges the receipt of FCT fellowship BPD/14398/2003. We
thank the anonymous referee for useful comments and suggestions.
\end{acknowledgements}
\bibliographystyle{aa}
\bibliography{flamare}

\Online

{\scriptsize\begin{longtable}{llrrrrllr}
\caption{\normalsize
Gas-phase oxygen abundances (computed with the $R_{\mathrm{23}}$
method) of star-forming galaxies at intermediate redshifts. \emph{LCL05}:
identification number, $^*$ flag for
candidate star-forming galaxies (see Paper I). \emph{alt}: alternative identification if available. 
\emph{z}: redshift. \emph{M(B)}:
absolute magnitude in the $B$ band (AB system). \emph{low:} lower
branch metallicity. \emph{high}: higher branch metallicity. \emph{N2}:
metallicity regime with the N2 diagnostic ({}``med'' stands for
galaxies with O/H$_{\mathrm{low}}$ greater than O/H$_{\mathrm{high}}$
which are kept as intermediate metallicity objects, see text for details).
\emph{LZ}: metallicity regime with the $L$ diagnostic (object with
an unreliable $L$ diagnostic are flagged by a `{*}'). \emph{12+log(O/H)}:
adopted abundance.}
\label{ohdata}\\
\hline\hline
LCL05	&alt	&z	&$M(B)$	&low	&high	&N2	&LZ	&12+log(O/H)	\\
\hline
\endfirsthead
\caption{\normalsize
continued}\\
\hline\hline
LCL05	&alt	&z	&$M(B)$	&low	&high	&N2	&LZ	&12+log(O/H)	\\
\hline
\endhead
\hline
\endfoot
\object{001}	&	&$0.341$	&$-19.47$	&$7.90$	&$8.79$	&...	&high	&$8.79\pm0.07$	\\
\object{002}	&	&$0.616$	&$-21.36$	&$7.38$	&$9.02$	&...	&high	&$9.02\pm0.03$	\\
\object{003}	&	&$0.341$	&$-20.31$	&$8.04$	&$8.61$	&...	&high	&$8.61\pm0.03$	\\
\object{004}	&\object{CFRS$\,$00.0852}	&$0.268$	&$-20.30$	&$8.02$	&$8.79$	&high	&high	&$8.79\pm0.05$	\\
\object{006}	&\object{CFRS$\,$00.0900}	&$0.247$	&$-18.85$	&$8.10$	&$8.56$	&low	&high$^\textrm{*}$	&$8.10\pm0.11$	\\
\object{007}	&\object{CFRS$\,$00.0940}	&$0.269$	&$-19.88$	&$8.39$	&$8.46$	&high	&high$^\textrm{*}$	&$8.46\pm0.28$	\\
\object{008}	&\object{CFRS$\,$00.1013}	&$0.244$	&$-19.13$	&$8.13$	&$8.58$	&high	&high$^\textrm{*}$	&$8.58\pm0.06$	\\
\object{009}	&\object{CFRS$\,$00.0124}	&$0.288$	&$-20.08$	&$7.96$	&$8.75$	&high	&high	&$8.75\pm0.06$	\\
\object{010}	&\object{CFRS$\,$00.0148}	&$0.267$	&$-18.93$	&$8.19$	&$8.60$	&high	&high$^\textrm{*}$	&$8.60\pm0.14$	\\
\object{011}	&\object{CFRS$\,$00.1726}	&$0.296$	&$-19.95$	&$8.58$	&$8.18$	&\textit{med}	&\textit{med}	&$8.38\pm0.30$	\\
\object{013}	&	&$0.249$	&$-18.73$	&$8.14$	&$8.64$	&...	&high$^\textrm{*}$	&$8.64\pm0.18$	\\
\object{014}	&	&$0.390$	&$-20.27$	&$7.82$	&$8.84$	&...	&high	&$8.84\pm0.08$	\\
\object{015}	&\object{CFRS$\,$00.1057}	&$0.243$	&$-17.31$	&$8.19$	&$8.47$	&high	&low$^\textrm{*}$	&$8.47\pm0.10$	\\
\object{016}	&\object{CFRS$\,$00.0121}	&$0.297$	&$-19.29$	&$8.26$	&$8.41$	&low	&high$^\textrm{*}$	&$8.26\pm0.03$	\\
\object{018}$^*$	&\object{CFRS$\,$03.1184}	&$0.205$	&$-18.46$	&$8.55$	&$8.19$	&\textit{med}	&\textit{med}	&$8.37\pm0.29$	\\
\object{020}	&\object{CFRS$\,$03.0442}	&$0.478$	&$-19.86$	&$8.04$	&$8.66$	&...	&high	&$8.66\pm0.25$	\\
\object{021}	&\object{CFRS$\,$03.0476}	&$0.260$	&$-18.87$	&$8.13$	&$8.55$	&...	&high$^\textrm{*}$	&$8.55\pm0.04$	\\
\object{022}	&\object{CFRS$\,$03.0488}	&$0.605$	&$-20.66$	&$7.88$	&$8.71$	&...	&high	&$8.71\pm0.06$	\\
\object{023}	&\object{CFRS$\,$03.0507}	&$0.465$	&$-20.86$	&$7.84$	&$8.77$	&...	&high	&$8.77\pm0.03$	\\
\object{024}	&\object{CFRS$\,$03.0523}	&$0.653$	&$-21.02$	&$7.93$	&$8.68$	&...	&high	&$8.68\pm0.05$	\\
\object{025}$^*$	&\object{CFRS$\,$03.0578}	&$0.219$	&$-19.18$	&$8.52$	&$8.21$	&\textit{med}	&\textit{med}	&$8.37\pm0.31$	\\
\object{026}	&\object{CFRS$\,$03.0605}	&$0.219$	&$-16.90$	&$8.09$	&$8.55$	&...	&low$^\textrm{*}$	&$8.09\pm0.08$	\\
\object{027}$^*$	&\object{CFRS$\,$03.0003}	&$0.219$	&$-17.48$	&$8.14$	&$8.50$	&...	&low$^\textrm{*}$	&$8.14\pm0.13$	\\
\object{028}	&\object{CFRS$\,$03.0037}	&$0.174$	&$-19.50$	&$8.23$	&$8.53$	&high	&high$^\textrm{*}$	&$8.53\pm0.03$	\\
\object{029}	&\object{CFRS$\,$03.0046}	&$0.512$	&$-20.88$	&$8.33$	&$8.43$	&...	&high	&$8.43\pm0.32$	\\
\object{030}	&\object{CFRS$\,$03.0085}	&$0.608$	&$-20.30$	&$7.71$	&$8.82$	&...	&high	&$8.82\pm0.16$	\\
\object{031}	&\object{CFRS$\,$03.0096}	&$0.219$	&$-17.58$	&$8.36$	&$8.34$	&\textit{med}	&\textit{med}	&$8.35\pm0.23$	\\
\object{032}	&\object{CFRS$\,$22.0502}	&$0.468$	&$-19.72$	&$7.89$	&$8.80$	&...	&high	&$8.80\pm0.06$	\\
\object{034}	&\object{CFRS$\,$22.0671}	&$0.318$	&$-19.20$	&$8.31$	&$8.42$	&low	&high$^\textrm{*}$	&$8.31\pm0.07$	\\
\object{035}	&\object{CFRS$\,$22.0819}	&$0.291$	&$-19.36$	&$7.89$	&$8.78$	&...	&high	&$8.78\pm0.05$	\\
\object{036}	&\object{CFRS$\,$22.0855}	&$0.210$	&$-18.77$	&$8.19$	&$8.49$	&high	&high$^\textrm{*}$	&$8.49\pm0.03$	\\
\object{038}	&\object{CFRS$\,$22.1013}	&$0.231$	&...	&$8.12$	&$8.57$	&high	&...	&$8.57\pm0.03$	\\
\object{039}	&\object{CFRS$\,$22.1084}	&$0.293$	&$-20.02$	&$8.09$	&$8.65$	&high	&high	&$8.65\pm0.06$	\\
\object{040}	&\object{CFRS$\,$22.1203}	&$0.538$	&$-20.58$	&$7.81$	&$8.76$	&...	&high	&$8.76\pm0.03$	\\
\object{041}	&	&$0.216$	&$-18.80$	&$8.63$	&$8.15$	&\textit{med}	&\textit{med}	&$8.39\pm0.31$	\\
\object{042}	&	&$0.352$	&$-18.77$	&$7.80$	&$8.73$	&...	&high	&$8.73\pm0.04$	\\
\object{043}	&\object{CFRS$\,$22.0622}	&$0.324$	&$-18.17$	&$7.95$	&$8.78$	&...	&high$^\textrm{*}$	&$8.78\pm0.10$	\\
\object{044}	&	&$0.276$	&$-19.80$	&$7.74$	&$8.87$	&...	&high	&$8.87\pm0.05$	\\
\object{046}	&	&$0.651$	&$-20.47$	&$7.64$	&$8.89$	&...	&high	&$8.89\pm0.04$	\\
\object{047}	&	&$0.472$	&$-18.35$	&$8.02$	&$8.62$	&...	&high$^\textrm{*}$	&$8.62\pm0.11$	\\
\object{049}	&\object{CFRS$\,$22.0832}	&$0.231$	&$-19.09$	&$7.79$	&$8.88$	&high	&high	&$8.88\pm0.04$	\\
\object{050}	&\object{CFRS$\,$22.1064}	&$0.537$	&$-19.96$	&$7.78$	&$8.77$	&...	&high	&$8.77\pm0.04$	\\
\object{051}$^*$	&\object{CFRS$\,$22.1339}	&$0.384$	&$-19.33$	&$7.97$	&$8.60$	&...	&high	&$8.60\pm0.02$	\\
\object{052}	&\object{CFRS$\,$22.0474}	&$0.279$	&$-18.62$	&$7.97$	&$8.61$	&high	&high$^\textrm{*}$	&$8.61\pm0.02$	\\
\object{053}	&\object{CFRS$\,$22.0504}	&$0.538$	&$-21.13$	&$7.86$	&$8.74$	&...	&high	&$8.74\pm0.03$	\\
\object{054}	&\object{CFRS$\,$22.0637}	&$0.542$	&$-21.19$	&$7.85$	&$8.74$	&...	&high	&$8.74\pm0.03$	\\
\object{055}	&\object{CFRS$\,$22.0642}	&$0.469$	&$-21.20$	&$7.92$	&$8.71$	&...	&high	&$8.71\pm0.05$	\\
\object{056}	&\object{CFRS$\,$22.0717}	&$0.279$	&$-20.15$	&$7.72$	&$8.94$	&high	&high	&$8.94\pm0.04$	\\
\object{057}	&\object{CFRS$\,$22.0823}	&$0.333$	&$-20.71$	&$8.18$	&$8.54$	&...	&high	&$8.54\pm0.09$	\\
\object{058}	&\object{CFRS$\,$22.1082}	&$0.292$	&$-19.83$	&$7.93$	&$8.62$	&...	&high	&$8.62\pm0.05$	\\
\object{059}	&	&$0.277$	&$-19.58$	&$7.99$	&$8.70$	&...	&high	&$8.70\pm0.05$	\\
\object{060}	&\object{CFRS$\,$22.1144}	&$0.359$	&$-19.23$	&$8.08$	&$8.60$	&...	&high$^\textrm{*}$	&$8.60\pm0.08$	\\
\object{061}	&\object{CFRS$\,$22.1220}	&$0.358$	&$-19.81$	&$8.14$	&$8.62$	&...	&high	&$8.62\pm0.04$	\\
\object{062}	&\object{CFRS$\,$22.1231}	&$0.285$	&$-19.22$	&$7.82$	&$8.85$	&high	&high	&$8.85\pm0.02$	\\
\object{063}	&\object{CFRS$\,$22.1309}	&$0.285$	&$-18.87$	&$8.03$	&$8.62$	&...	&high$^\textrm{*}$	&$8.62\pm0.05$	\\
\object{065}	&	&$0.066$	&$-15.41$	&$8.19$	&$8.46$	&low	&low	&$8.19\pm0.01$	\\
\object{067}	&	&$0.450$	&$-19.18$	&$8.15$	&$8.58$	&...	&high$^\textrm{*}$	&$8.58\pm0.12$	\\
\object{068}	&	&$0.629$	&$-19.70$	&$8.54$	&$8.20$	&\textit{med}	&\textit{med}	&$8.37\pm0.63$	\\
\object{069}	&	&$0.221$	&$-18.43$	&$8.23$	&$8.49$	&high	&high$^\textrm{*}$	&$8.49\pm0.07$	\\
\object{070}	&	&$0.739$	&$-20.39$	&$7.57$	&$8.94$	&...	&high	&$8.94\pm0.02$	\\
\object{071}	&	&$0.526$	&$-20.28$	&$8.15$	&$8.58$	&...	&high	&$8.58\pm0.04$	\\
\object{072}	&	&$0.342$	&$-18.92$	&$7.93$	&$8.74$	&...	&high	&$8.74\pm0.06$	\\
\object{073}	&\object{LBP2003 b}	&$0.260$	&$-19.44$	&$7.85$	&$8.84$	&high	&high	&$8.84\pm0.02$	\\
\object{076}	&\object{LBP2003 h}	&$0.321$	&$-20.40$	&$7.96$	&$8.77$	&high	&high	&$8.77\pm0.03$	\\
\object{077}	&\object{LBP2003 c}	&$0.300$	&$-19.87$	&$7.74$	&$8.90$	&high	&high	&$8.90\pm0.03$	\\
\object{078}	&\object{CPK2001 V7}	&$0.567$	&$-20.97$	&$7.92$	&$8.75$	&...	&high	&$8.75\pm0.06$	\\
\object{079}	&\object{CPK2001 V6}	&$0.410$	&$-19.16$	&$8.08$	&$8.68$	&...	&high$^\textrm{*}$	&$8.68\pm0.11$	\\
\object{080}	&\object{CBB2001 688}	&$0.330$	&$-18.82$	&$7.90$	&$8.73$	&high	&high	&$8.73\pm0.03$	\\
\object{081}	&\object{CPK2001 V11}	&$0.380$	&$-17.53$	&$7.80$	&$8.76$	&...	&low$^\textrm{*}$	&$7.80\pm0.15$	\\
\object{083}	&	&$0.726$	&$-20.05$	&$8.14$	&$8.51$	&...	&high	&$8.51\pm0.04$	\\
\object{084}	&\object{CBB2001 453}	&$0.410$	&$-18.82$	&$8.62$	&$8.12$	&\textit{med}	&\textit{med}	&$8.37\pm0.26$	\\
\object{085}$^*$	&	&$0.719$	&$-19.59$	&$8.47$	&$8.26$	&\textit{med}	&\textit{med}	&$8.36\pm0.52$	\\
\object{086}	&	&$0.412$	&$-19.95$	&$8.23$	&$8.48$	&...	&high$^\textrm{*}$	&$8.48\pm0.11$	\\
\object{087}	&	&$0.409$	&$-18.32$	&$7.86$	&$8.70$	&...	&high$^\textrm{*}$	&$8.70\pm0.06$	\\
\object{088}$^*$	&	&$0.757$	&$-19.78$	&$8.08$	&$8.52$	&...	&high	&$8.52\pm0.05$	\\
\object{089}$^*$	&	&$0.563$	&$-20.02$	&$8.71$	&$8.02$	&\textit{med}	&\textit{med}	&$8.37\pm0.36$	\\
\object{091}	&	&$0.786$	&$-19.41$	&$7.83$	&$8.74$	&...	&high	&$8.74\pm0.06$	\\
\object{092}$^*$	&	&$0.329$	&$-18.43$	&$8.09$	&$8.53$	&...	&high$^\textrm{*}$	&$8.53\pm0.02$	\\
\object{093}	&	&$0.640$	&$-20.33$	&$7.98$	&$8.64$	&...	&high	&$8.64\pm0.06$	\\
\object{095}	&	&$0.342$	&$-17.87$	&$7.47$	&$8.95$	&...	&high	&$8.95\pm0.02$	\\
\object{096}	&	&$0.439$	&$-19.37$	&$8.40$	&$8.37$	&\textit{med}	&\textit{med}	&$8.39\pm0.28$	\\
\object{098}	&	&$0.572$	&$-17.62$	&$8.34$	&$8.37$	&...	&low$^\textrm{*}$	&$8.34\pm0.32$	\\
\object{099}	&	&$0.355$	&$-19.93$	&$7.95$	&$8.74$	&...	&high	&$8.74\pm0.05$	\\
\object{100}	&	&$0.355$	&$-19.40$	&$8.01$	&$8.65$	&...	&high	&$8.65\pm0.05$	\\
\object{101}	&	&$0.355$	&$-19.50$	&$8.56$	&$8.23$	&\textit{med}	&\textit{med}	&$8.39\pm0.35$	\\
\object{102}	&	&$0.428$	&$-21.01$	&$8.59$	&$8.30$	&\textit{med}	&\textit{med}	&$8.45\pm0.28$	\\
\object{103}	&	&$0.476$	&$-21.26$	&$8.23$	&$8.52$	&...	&high	&$8.52\pm0.07$	\\
\object{104}	&	&$0.452$	&$-20.04$	&$7.98$	&$8.78$	&...	&high	&$8.78\pm0.07$	\\
\object{105}$^*$	&\object{GDDS$\,$02-0452}	&$0.828$	&$-20.79$	&$8.75$	&$7.99$	&\textit{med}	&\textit{med}	&$8.37\pm0.43$	\\
\object{106}	&\object{GDDS$\,$02-0585}	&$0.825$	&$-19.70$	&$8.23$	&$8.46$	&...	&high$^\textrm{*}$	&$8.46\pm0.17$	\\
\object{107}	&\object{GDDS$\,$02-0756}	&$0.864$	&$-18.89$	&$8.57$	&$8.20$	&\textit{med}	&\textit{med}	&$8.38\pm0.37$	\\
\object{108}	&\object{GDDS$\,$02-0995}	&$0.786$	&$-19.17$	&$8.55$	&$8.23$	&\textit{med}	&\textit{med}	&$8.39\pm0.53$	\\
\object{110}	&\object{GDDS$\,$02-1724}	&$0.996$	&$-22.08$	&$7.88$	&$8.65$	&...	&high	&$8.65\pm0.16$	\\
\object{112}	&\object{GDDS$\,$12-5513}	&$0.611$	&$-20.43$	&$7.86$	&$8.84$	&...	&high	&$8.84\pm0.03$	\\
\object{113}	&\object{GDDS$\,$12-5685}	&$0.960$	&$-20.68$	&$7.84$	&$8.74$	&...	&high	&$8.74\pm0.10$	\\
\object{114}	&\object{GDDS$\,$12-5722}	&$0.841$	&$-20.28$	&$7.90$	&$8.75$	&...	&high	&$8.75\pm0.07$	\\
\object{116}	&\object{GDDS$\,$12-6800}	&$0.615$	&$-21.60$	&$7.92$	&$8.72$	&...	&high	&$8.72\pm0.13$	\\
\object{117}	&\object{GDDS$\,$12-7099}	&$0.567$	&$-20.09$	&$7.60$	&$9.03$	&...	&high	&$9.03\pm0.01$	\\
\object{118}	&\object{GDDS$\,$12-7205}	&$0.568$	&$-19.54$	&$7.41$	&$9.02$	&...	&high	&$9.02\pm0.01$	\\
\object{119}	&\object{GDDS$\,$12-7660}	&$0.791$	&$-20.49$	&$7.81$	&$8.83$	&...	&high	&$8.83\pm0.03$	\\
\object{120}	&\object{GDDS$\,$12-7939}	&$0.664$	&$-18.95$	&$8.13$	&$8.53$	&...	&high$^\textrm{*}$	&$8.53\pm0.09$	\\
\object{122}	&\object{GDDS$\,$22-0040}	&$0.818$	&$-19.73$	&$7.97$	&$8.63$	&...	&high	&$8.63\pm0.07$	\\
\object{123}	&\object{GDDS$\,$22-0145}	&$0.754$	&$-19.43$	&$7.91$	&$8.69$	&...	&high	&$8.69\pm0.04$	\\
\object{124}	&\object{GDDS$\,$22-0563}	&$0.787$	&$-20.37$	&$7.82$	&$8.78$	&...	&high	&$8.78\pm0.02$	\\
\object{125}	&\object{GDDS$\,$22-0619}	&$0.673$	&$-18.68$	&$8.22$	&$8.54$	&...	&high$^\textrm{*}$	&$8.54\pm0.22$	\\
\object{126}	&\object{GDDS$\,$22-0630}	&$0.753$	&$-20.18$	&$7.24$	&$9.03$	&...	&high	&$9.03\pm0.01$	\\
\object{127}	&\object{GDDS$\,$22-0643}	&$0.788$	&$-19.56$	&$8.11$	&$8.59$	&...	&high	&$8.59\pm0.14$	\\
\object{129}	&\object{GDDS$\,$22-0926}	&$0.786$	&$-19.30$	&$8.19$	&$8.49$	&...	&high$^\textrm{*}$	&$8.49\pm0.14$	\\
\object{131}	&\object{GDDS$\,$22-1674}	&$0.879$	&$-19.23$	&$7.70$	&$8.77$	&...	&high	&$8.77\pm0.03$	\\
\object{132}	&\object{GDDS$\,$22-2196}	&$0.627$	&$-18.30$	&$8.30$	&$8.40$	&...	&high$^\textrm{*}$	&$8.40\pm0.18$	\\
\object{133}	&\object{GDDS$\,$22-2491}	&$0.471$	&$-17.69$	&$8.05$	&$8.56$	&low	&low$^\textrm{*}$	&$8.05\pm0.08$	\\
\object{134}	&\object{GDDS$\,$22-2541}	&$0.617$	&$-19.69$	&$7.96$	&$8.71$	&...	&high	&$8.71\pm0.06$	\\
\object{135}	&\object{GDDS$\,$22-2639}	&$0.883$	&$-20.91$	&$7.54$	&$8.88$	&...	&high	&$8.88\pm0.01$	\\
\object{136}	&\object{SKK2001 368}	&$0.693$	&$-20.91$	&$7.64$	&$8.98$	&...	&high	&$8.98\pm0.04$	\\
\object{137}	&\object{SKK2001 159}	&$0.473$	&$-19.62$	&$7.75$	&$8.94$	&...	&high	&$8.94\pm0.04$	\\
\object{138}	&	&$0.449$	&$-17.77$	&$8.13$	&$8.60$	&...	&low$^\textrm{*}$	&$8.13\pm0.19$	\\
\object{140}$^*$	&	&$0.731$	&$-19.61$	&$8.40$	&$8.32$	&\textit{med}	&\textit{med}	&$8.36\pm0.36$	\\
\object{141}	&	&$0.398$	&$-19.09$	&$8.88$	&$7.83$	&\textit{med}	&\textit{med}	&$8.35\pm0.81$	\\
\end{longtable}}

{\scriptsize\begin{longtable}{llrrrrrr}
\caption{\normalsize
Gas-phase oxygen abundances (with the CL01 method) of star-forming
galaxies at intermediate redshifts. \emph{LCL05}: identification number, $^*$ flag for
candidate star-forming galaxies (see Paper I). 
\emph{alt}: alternative identification if available. \emph{low:} lower peak metallicity. \emph{$A_{l}$}:
associated likelihood amplitude. \emph{$P_{l}$}: associated probability.
\emph{high}: higher peak metallicity. \emph{$A_{h}$}: associated
likelihood amplitude. \emph{$P_{h}$}: associated probability. The
boldface values designate the highest likelihood and the adopted abundances.}
\label{ohdatacl}\\
\hline\hline
LCL05	&alt	&low	&$A_l$	&$P_l$	&high	&$A_h$	&$P_h$	\\
\hline
\endfirsthead
\caption{\normalsize
continued}\\
\hline\hline
LCL05	&alt	&low	&$A_l$	&$P_l$	&high	&$A_h$	&$P_h$	\\
\hline
\endhead
\hline
\endfoot
\object{001}	&	&$7.80\pm0.08$	&$0.33$	&$0.03$	&$\mathbf{8.88\pm0.10}$	&$\mathbf{9.64}$	&$\mathbf{0.97}$	\\
\object{002}	&	&$7.75\pm0.07$	&$0.01$	&$0.00$	&$\mathbf{9.10\pm0.09}$	&$\mathbf{10.56}$	&$\mathbf{1.00}$	\\
\object{003}	&	&$8.21\pm0.21$	&$2.78$	&$0.45$	&$\mathbf{8.65\pm0.11}$	&$\mathbf{3.35}$	&$\mathbf{0.55}$	\\
\object{004}	&\object{CFRS$\,$00.0852}	&$8.30\pm0.35$	&$0.02$	&$0.00$	&$\mathbf{8.79\pm0.02}$	&$\mathbf{33.43}$	&$\mathbf{1.00}$	\\
\object{006}	&\object{CFRS$\,$00.0900}	&$\mathbf{8.06\pm0.08}$	&$\mathbf{6.65}$	&$\mathbf{0.72}$	&$8.28\pm0.17$	&$2.63$	&$0.28$	\\
\object{007}	&\object{CFRS$\,$00.0940}	&$7.97\pm0.12$	&$1.33$	&$0.12$	&$\mathbf{8.82\pm0.08}$	&$\mathbf{9.76}$	&$\mathbf{0.88}$	\\
\object{008}	&\object{CFRS$\,$00.1013}	&$8.30\pm0.15$	&$2.23$	&$0.20$	&$\mathbf{8.67\pm0.07}$	&$\mathbf{8.98}$	&$\mathbf{0.80}$	\\
\object{009}	&\object{CFRS$\,$00.0124}	&$8.81\pm0.07$	&$5.16$	&$0.26$	&$\mathbf{8.88\pm0.04}$	&$\mathbf{14.95}$	&$\mathbf{0.74}$	\\
\object{010}	&\object{CFRS$\,$00.0148}	&$8.75\pm0.01$	&$1.65$	&$0.11$	&$\mathbf{8.94\pm0.07}$	&$\mathbf{13.73}$	&$\mathbf{0.89}$	\\
\object{011}	&\object{CFRS$\,$00.1726}	&$\mathbf{8.41\pm0.09}$	&$\mathbf{7.34}$	&$\mathbf{0.53}$	&$8.44\pm0.06$	&$6.40$	&$0.47$	\\
\object{012}	&\object{CFRS$\,$00.0699}	&$\mathbf{8.38\pm0.29}$	&$\mathbf{3.11}$	&$\mathbf{0.70}$	&$8.44\pm0.07$	&$1.33$	&$0.30$	\\
\object{013}	&	&$8.26\pm0.31$	&$0.89$	&$0.14$	&$\mathbf{8.75\pm0.13}$	&$\mathbf{5.70}$	&$\mathbf{0.86}$	\\
\object{014}	&	&$7.80\pm0.09$	&$0.23$	&$0.02$	&$\mathbf{9.04\pm0.10}$	&$\mathbf{9.13}$	&$\mathbf{0.98}$	\\
\object{015}	&\object{CFRS$\,$00.1057}	&$8.18\pm0.15$	&$3.24$	&$0.47$	&$\mathbf{8.58\pm0.13}$	&$\mathbf{3.70}$	&$\mathbf{0.53}$	\\
\object{016}	&\object{CFRS$\,$00.0121}	&$8.22\pm0.02$	&$13.40$	&$0.31$	&$\mathbf{8.41\pm0.02}$	&$\mathbf{29.70}$	&$\mathbf{0.69}$	\\
\object{018}$^*$	&\object{CFRS$\,$03.1184}	&$8.19\pm0.06$	&$1.32$	&$0.15$	&$\mathbf{8.44\pm0.12}$	&$\mathbf{7.34}$	&$\mathbf{0.85}$	\\
\object{019}	&\object{CFRS$\,$03.1343}	&$\mathbf{8.11\pm0.12}$	&$\mathbf{6.37}$	&$\mathbf{0.82}$	&$8.36\pm0.14$	&$1.44$	&$0.18$	\\
\object{020}	&\object{CFRS$\,$03.0442}	&$8.19\pm0.24$	&$2.29$	&$0.46$	&$\mathbf{8.69\pm0.17}$	&$\mathbf{2.65}$	&$\mathbf{0.54}$	\\
\object{021}	&\object{CFRS$\,$03.0476}	&$8.05\pm0.08$	&$6.16$	&$0.21$	&$\mathbf{8.06\pm0.02}$	&$\mathbf{23.80}$	&$\mathbf{0.79}$	\\
\object{022}	&\object{CFRS$\,$03.0488}	&$8.20\pm0.21$	&$2.79$	&$0.47$	&$\mathbf{8.65\pm0.13}$	&$\mathbf{3.16}$	&$\mathbf{0.53}$	\\
\object{023}	&\object{CFRS$\,$03.0507}	&$8.21\pm0.30$	&$2.29$	&$0.47$	&$\mathbf{8.80\pm0.12}$	&$\mathbf{2.54}$	&$\mathbf{0.53}$	\\
\object{024}	&\object{CFRS$\,$03.0523}	&$8.16\pm0.24$	&$2.30$	&$0.44$	&$\mathbf{8.70\pm0.15}$	&$\mathbf{2.90}$	&$\mathbf{0.56}$	\\
\object{025}$^*$	&\object{CFRS$\,$03.0578}	&$8.44\pm0.13$	&$4.47$	&$0.42$	&$\mathbf{8.47\pm0.07}$	&$\mathbf{6.13}$	&$\mathbf{0.58}$	\\
\object{026}	&\object{CFRS$\,$03.0605}	&$\mathbf{8.30\pm0.18}$	&$\mathbf{4.14}$	&$\mathbf{0.68}$	&$8.60\pm0.12$	&$1.99$	&$0.32$	\\
\object{027}$^*$	&\object{CFRS$\,$03.0003}	&$8.07\pm0.10$	&$1.20$	&$0.26$	&$\mathbf{8.42\pm0.25}$	&$\mathbf{3.43}$	&$\mathbf{0.74}$	\\
\object{028}	&\object{CFRS$\,$03.0037}	&$\mathbf{8.74\pm0.02}$	&$\mathbf{46.04}$	&$\mathbf{0.93}$	&$8.81\pm0.01$	&$3.64$	&$0.07$	\\
\object{029}	&\object{CFRS$\,$03.0046}	&$8.23\pm0.21$	&$2.64$	&$0.49$	&$\mathbf{8.62\pm0.16}$	&$\mathbf{2.79}$	&$\mathbf{0.51}$	\\
\object{030}	&\object{CFRS$\,$03.0085}	&$8.15\pm0.25$	&$1.91$	&$0.40$	&$\mathbf{8.74\pm0.18}$	&$\mathbf{2.88}$	&$\mathbf{0.60}$	\\
\object{031}	&\object{CFRS$\,$03.0096}	&$\mathbf{8.36\pm0.15}$	&$\mathbf{6.48}$	&$\mathbf{1.00}$	&$9.15\pm0.02$	&$0.00$	&$0.00$	\\
\object{032}	&\object{CFRS$\,$22.0502}	&$7.85\pm0.08$	&$1.58$	&$0.17$	&$\mathbf{8.81\pm0.11}$	&$\mathbf{7.47}$	&$\mathbf{0.83}$	\\
\object{033}	&\object{CFRS$\,$22.0585}	&..	&..	&..	&$\mathbf{9.30\pm0.09}$	&$\mathbf{11.14}$	&$\mathbf{1.00}$	\\
\object{034}	&\object{CFRS$\,$22.0671}	&$\mathbf{8.35\pm0.15}$	&$\mathbf{6.33}$	&$\mathbf{1.00}$	&..	&..	&..	\\
\object{035}	&\object{CFRS$\,$22.0819}	&$7.86\pm0.06$	&$3.10$	&$0.30$	&$\mathbf{8.76\pm0.11}$	&$\mathbf{7.34}$	&$\mathbf{0.70}$	\\
\object{036}	&\object{CFRS$\,$22.0855}	&$8.50\pm0.17$	&$0.24$	&$0.01$	&$\mathbf{8.68\pm0.02}$	&$\mathbf{46.82}$	&$\mathbf{0.99}$	\\
\object{037}$^*$	&\object{CFRS$\,$22.0975}	&..	&..	&..	&$\mathbf{8.53\pm0.16}$	&$\mathbf{6.16}$	&$\mathbf{1.00}$	\\
\object{038}	&\object{CFRS$\,$22.1013}	&$8.21\pm0.08$	&$0.99$	&$0.10$	&$\mathbf{8.50\pm0.07}$	&$\mathbf{9.31}$	&$\mathbf{0.90}$	\\
\object{039}	&\object{CFRS$\,$22.1084}	&$8.76\pm0.05$	&$9.11$	&$0.37$	&$\mathbf{8.83\pm0.03}$	&$\mathbf{15.48}$	&$\mathbf{0.63}$	\\
\object{040}	&\object{CFRS$\,$22.1203}	&$8.27\pm0.30$	&$2.42$	&$0.47$	&$\mathbf{8.77\pm0.09}$	&$\mathbf{2.77}$	&$\mathbf{0.53}$	\\
\object{041}	&	&$8.19\pm0.24$	&$2.42$	&$0.48$	&$\mathbf{8.72\pm0.15}$	&$\mathbf{2.59}$	&$\mathbf{0.52}$	\\
\object{042}	&	&$8.07\pm0.12$	&$2.49$	&$0.43$	&$\mathbf{8.51\pm0.21}$	&$\mathbf{3.29}$	&$\mathbf{0.57}$	\\
\object{043}	&\object{CFRS$\,$22.0622}	&$7.87\pm0.11$	&$1.02$	&$0.13$	&$\mathbf{8.82\pm0.12}$	&$\mathbf{6.89}$	&$\mathbf{0.87}$	\\
\object{044}	&	&$7.80\pm0.08$	&$1.04$	&$0.12$	&$\mathbf{8.93\pm0.12}$	&$\mathbf{7.56}$	&$\mathbf{0.88}$	\\
\object{046}	&	&$7.86\pm0.09$	&$3.00$	&$0.36$	&$\mathbf{8.98\pm0.13}$	&$\mathbf{5.26}$	&$\mathbf{0.64}$	\\
\object{047}	&	&$8.14\pm0.14$	&$2.31$	&$0.38$	&$\mathbf{8.51\pm0.18}$	&$\mathbf{3.74}$	&$\mathbf{0.62}$	\\
\object{049}	&\object{CFRS$\,$22.0832}	&..	&..	&..	&$\mathbf{8.95\pm0.02}$	&$\mathbf{41.23}$	&$\mathbf{1.00}$	\\
\object{050}	&\object{CFRS$\,$22.1064}	&$8.07\pm0.22$	&$2.16$	&$0.41$	&$\mathbf{8.80\pm0.16}$	&$\mathbf{3.12}$	&$\mathbf{0.59}$	\\
\object{051}$^*$	&\object{CFRS$\,$22.1339}	&$8.11\pm0.11$	&$2.36$	&$0.42$	&$\mathbf{8.52\pm0.21}$	&$\mathbf{3.25}$	&$\mathbf{0.58}$	\\
\object{052}	&\object{CFRS$\,$22.0474}	&$8.98\pm0.01$	&$9.07$	&$0.17$	&$\mathbf{9.08\pm0.01}$	&$\mathbf{45.79}$	&$\mathbf{0.83}$	\\
\object{053}	&\object{CFRS$\,$22.0504}	&$8.27\pm0.30$	&$2.45$	&$0.47$	&$\mathbf{8.78\pm0.09}$	&$\mathbf{2.80}$	&$\mathbf{0.53}$	\\
\object{054}	&\object{CFRS$\,$22.0637}	&$8.27\pm0.30$	&$2.45$	&$0.46$	&$\mathbf{8.77\pm0.09}$	&$\mathbf{2.82}$	&$\mathbf{0.54}$	\\
\object{055}	&\object{CFRS$\,$22.0642}	&$\mathbf{8.67\pm0.11}$	&$\mathbf{8.66}$	&$\mathbf{0.98}$	&$9.39\pm0.05$	&$0.14$	&$0.02$	\\
\object{056}	&\object{CFRS$\,$22.0717}	&$9.00\pm0.35$	&$0.16$	&$0.01$	&$\mathbf{9.03\pm0.06}$	&$\mathbf{14.07}$	&$\mathbf{0.99}$	\\
\object{057}	&\object{CFRS$\,$22.0823}	&$8.04\pm0.20$	&$2.21$	&$0.41$	&$\mathbf{8.83\pm0.17}$	&$\mathbf{3.17}$	&$\mathbf{0.59}$	\\
\object{058}	&\object{CFRS$\,$22.1082}	&$\mathbf{8.41\pm0.10}$	&$\mathbf{6.60}$	&$\mathbf{0.87}$	&$8.49\pm0.32$	&$0.98$	&$0.13$	\\
\object{059}	&	&$7.97\pm0.12$	&$3.06$	&$0.43$	&$\mathbf{8.70\pm0.15}$	&$\mathbf{4.07}$	&$\mathbf{0.57}$	\\
\object{060}	&\object{CFRS$\,$22.1144}	&$8.24\pm0.17$	&$3.26$	&$0.50$	&$\mathbf{8.60\pm0.14}$	&$\mathbf{3.29}$	&$\mathbf{0.50}$	\\
\object{061}	&\object{CFRS$\,$22.1220}	&$7.95\pm0.08$	&$3.86$	&$0.43$	&$\mathbf{8.70\pm0.13}$	&$\mathbf{5.13}$	&$\mathbf{0.57}$	\\
\object{062}	&\object{CFRS$\,$22.1231}	&..	&..	&..	&$\mathbf{9.39\pm0.03}$	&$\mathbf{34.27}$	&$\mathbf{1.00}$	\\
\object{063}	&\object{CFRS$\,$22.1309}	&$8.27\pm0.23$	&$3.00$	&$0.50$	&$\mathbf{8.69\pm0.10}$	&$\mathbf{3.02}$	&$\mathbf{0.50}$	\\
\object{065}	&	&$\mathbf{8.07\pm0.01}$	&$\mathbf{50.13}$	&$\mathbf{1.00}$	&..	&..	&..	\\
\object{066}	&	&$8.08\pm0.16$	&$2.19$	&$0.42$	&$\mathbf{8.56\pm0.21}$	&$\mathbf{3.02}$	&$\mathbf{0.58}$	\\
\object{067}	&	&$\mathbf{8.24\pm0.18}$	&$\mathbf{3.05}$	&$\mathbf{0.51}$	&$8.60\pm0.15$	&$2.96$	&$0.49$	\\
\object{068}	&	&$\mathbf{8.26\pm0.22}$	&$\mathbf{2.65}$	&$\mathbf{0.51}$	&$8.60\pm0.17$	&$2.59$	&$0.49$	\\
\object{069}	&	&$\mathbf{8.29\pm0.15}$	&$\mathbf{5.64}$	&$\mathbf{0.69}$	&$8.61\pm0.05$	&$2.58$	&$0.31$	\\
\object{070}	&	&$7.78\pm0.06$	&$1.84$	&$0.21$	&$\mathbf{8.92\pm0.12}$	&$\mathbf{7.12}$	&$\mathbf{0.79}$	\\
\object{071}	&	&$\mathbf{8.40\pm0.18}$	&$\mathbf{5.45}$	&$\mathbf{1.00}$	&..	&..	&..	\\
\object{072}	&	&$7.97\pm0.15$	&$2.52$	&$0.41$	&$\mathbf{8.71\pm0.17}$	&$\mathbf{3.66}$	&$\mathbf{0.59}$	\\
\object{073}	&\object{LBP2003 b}	&..	&..	&..	&$\mathbf{8.91\pm0.02}$	&$\mathbf{62.33}$	&$\mathbf{1.00}$	\\
\object{074}	&	&$\mathbf{8.12\pm0.01}$	&$\mathbf{50.13}$	&$\mathbf{1.00}$	&..	&..	&..	\\
\object{075}	&\object{CBB2001 796}	&$\mathbf{8.06\pm0.10}$	&$\mathbf{9.70}$	&$\mathbf{1.00}$	&..	&..	&..	\\
\object{076}	&\object{LBP2003 h}	&$\mathbf{8.89\pm0.01}$	&$\mathbf{167.13}$	&$\mathbf{1.00}$	&$8.96\pm0.35$	&$0.09$	&$0.00$	\\
\object{077}	&\object{LBP2003 c}	&..	&..	&..	&$\mathbf{9.05\pm0.02}$	&$\mathbf{46.32}$	&$\mathbf{1.00}$	\\
\object{078}	&\object{CPK2001 V7}	&$7.96\pm0.17$	&$2.19$	&$0.36$	&$\mathbf{8.84\pm0.15}$	&$\mathbf{3.93}$	&$\mathbf{0.64}$	\\
\object{079}	&\object{CPK2001 V6}	&$7.92\pm0.14$	&$1.64$	&$0.23$	&$\mathbf{8.78\pm0.13}$	&$\mathbf{5.50}$	&$\mathbf{0.77}$	\\
\object{080}	&\object{CBB2001 688}	&$7.95\pm0.05$	&$3.94$	&$0.21$	&$\mathbf{8.79\pm0.05}$	&$\mathbf{15.02}$	&$\mathbf{0.79}$	\\
\object{081}	&\object{CPK2001 V11}	&$8.12\pm0.23$	&$2.24$	&$0.45$	&$\mathbf{8.74\pm0.18}$	&$\mathbf{2.70}$	&$\mathbf{0.55}$	\\
\object{083}	&	&$\mathbf{8.38\pm0.14}$	&$\mathbf{6.62}$	&$\mathbf{0.87}$	&$8.76\pm0.05$	&$0.98$	&$0.13$	\\
\object{084}	&\object{CBB2001 453}	&..	&..	&..	&$\mathbf{8.75\pm0.05}$	&$\mathbf{17.46}$	&$\mathbf{1.00}$	\\
\object{085}$^*$	&	&$\mathbf{8.30\pm0.19}$	&$\mathbf{3.59}$	&$\mathbf{0.62}$	&$8.60\pm0.14$	&$2.23$	&$0.38$	\\
\object{086}	&	&$8.12\pm0.19$	&$2.33$	&$0.44$	&$\mathbf{8.60\pm0.19}$	&$\mathbf{2.97}$	&$\mathbf{0.56}$	\\
\object{087}	&	&$8.15\pm0.24$	&$2.27$	&$0.45$	&$\mathbf{8.70\pm0.16}$	&$\mathbf{2.83}$	&$\mathbf{0.55}$	\\
\object{088}$^*$	&	&$\mathbf{8.36\pm0.19}$	&$\mathbf{3.98}$	&$\mathbf{0.66}$	&$8.44\pm0.06$	&$2.01$	&$0.34$	\\
\object{089}$^*$	&	&$\mathbf{8.44\pm0.14}$	&$\mathbf{5.25}$	&$\mathbf{0.59}$	&$8.48\pm0.07$	&$3.62$	&$0.41$	\\
\object{090}$^*$	&	&$\mathbf{8.45\pm0.01}$	&$\mathbf{51.09}$	&$\mathbf{0.83}$	&$8.52\pm0.08$	&$10.39$	&$0.17$	\\
\object{091}	&	&$8.09\pm0.22$	&$2.19$	&$0.43$	&$\mathbf{8.80\pm0.17}$	&$\mathbf{2.94}$	&$\mathbf{0.57}$	\\
\object{092}$^*$	&	&..	&..	&..	&$\mathbf{8.39\pm0.18}$	&$\mathbf{5.39}$	&$\mathbf{1.00}$	\\
\object{093}	&	&$8.20\pm0.20$	&$2.79$	&$0.47$	&$\mathbf{8.64\pm0.14}$	&$\mathbf{3.16}$	&$\mathbf{0.53}$	\\
\object{095}	&	&$7.81\pm0.06$	&$2.29$	&$0.26$	&$\mathbf{9.05\pm0.13}$	&$\mathbf{6.38}$	&$\mathbf{0.74}$	\\
\object{096}	&	&$\mathbf{8.43\pm0.17}$	&$\mathbf{6.02}$	&$\mathbf{1.00}$	&..	&..	&..	\\
\object{098}	&	&$\mathbf{8.28\pm0.20}$	&$\mathbf{2.96}$	&$\mathbf{0.53}$	&$8.60\pm0.15$	&$2.67$	&$0.47$	\\
\object{099}	&	&$7.88\pm0.09$	&$3.23$	&$0.38$	&$\mathbf{8.77\pm0.13}$	&$\mathbf{5.19}$	&$\mathbf{0.62}$	\\
\object{100}	&	&$8.26\pm0.26$	&$2.77$	&$0.48$	&$\mathbf{8.71\pm0.10}$	&$\mathbf{2.95}$	&$\mathbf{0.52}$	\\
\object{101}	&	&$\mathbf{8.42\pm0.20}$	&$\mathbf{4.95}$	&$\mathbf{1.00}$	&..	&..	&..	\\
\object{102}	&	&$8.76\pm0.10$	&$9.71$	&$0.37$	&$\mathbf{8.85\pm0.01}$	&$\mathbf{16.31}$	&$\mathbf{0.63}$	\\
\object{103}	&	&$\mathbf{8.25\pm0.16}$	&$\mathbf{3.67}$	&$\mathbf{0.54}$	&$8.60\pm0.13$	&$3.19$	&$0.46$	\\
\object{104}	&	&$7.83\pm0.09$	&$1.08$	&$0.13$	&$\mathbf{8.91\pm0.11}$	&$\mathbf{7.38}$	&$\mathbf{0.87}$	\\
\object{105}$^*$	&\object{GDDS$\,$02-0452}	&$8.18\pm0.20$	&$2.37$	&$0.46$	&$\mathbf{8.60\pm0.19}$	&$\mathbf{2.78}$	&$\mathbf{0.54}$	\\
\object{106}	&\object{GDDS$\,$02-0585}	&$8.07\pm0.13$	&$1.41$	&$0.29$	&$\mathbf{8.46\pm0.24}$	&$\mathbf{3.40}$	&$\mathbf{0.71}$	\\
\object{107}	&\object{GDDS$\,$02-0756}	&$\mathbf{8.30\pm0.19}$	&$\mathbf{3.38}$	&$\mathbf{0.57}$	&$8.60\pm0.14$	&$2.53$	&$0.43$	\\
\object{108}	&\object{GDDS$\,$02-0995}	&$8.08\pm0.22$	&$1.72$	&$0.40$	&$\mathbf{8.78\pm0.24}$	&$\mathbf{2.56}$	&$\mathbf{0.60}$	\\
\object{110}	&\object{GDDS$\,$02-1724}	&$8.05\pm0.27$	&$2.01$	&$0.37$	&$\mathbf{8.85\pm0.12}$	&$\mathbf{3.41}$	&$\mathbf{0.63}$	\\
\object{111}$^*$	&\object{GDDS$\,$12-5337}	&$\mathbf{8.48\pm0.12}$	&$\mathbf{7.60}$	&$\mathbf{0.77}$	&$8.49\pm0.03$	&$2.27$	&$0.23$	\\
\object{112}	&\object{GDDS$\,$12-5513}	&$7.69\pm0.02$	&$0.02$	&$0.00$	&$\mathbf{9.04\pm0.10}$	&$\mathbf{9.77}$	&$\mathbf{1.00}$	\\
\object{113}	&\object{GDDS$\,$12-5685}	&$8.04\pm0.18$	&$2.15$	&$0.47$	&$\mathbf{8.71\pm0.25}$	&$\mathbf{2.41}$	&$\mathbf{0.53}$	\\
\object{114}	&\object{GDDS$\,$12-5722}	&$7.87\pm0.10$	&$2.52$	&$0.35$	&$\mathbf{8.99\pm0.15}$	&$\mathbf{4.66}$	&$\mathbf{0.65}$	\\
\object{116}	&\object{GDDS$\,$12-6800}	&$7.88\pm0.13$	&$1.48$	&$0.24$	&$\mathbf{9.03\pm0.17}$	&$\mathbf{4.58}$	&$\mathbf{0.76}$	\\
\object{117}	&\object{GDDS$\,$12-7099}	&..	&..	&..	&$\mathbf{9.19\pm0.05}$	&$\mathbf{15.83}$	&$\mathbf{1.00}$	\\
\object{118}	&\object{GDDS$\,$12-7205}	&..	&..	&..	&$\mathbf{9.13\pm0.08}$	&$\mathbf{12.13}$	&$\mathbf{1.00}$	\\
\object{119}	&\object{GDDS$\,$12-7660}	&$7.77\pm0.07$	&$0.68$	&$0.07$	&$\mathbf{8.99\pm0.11}$	&$\mathbf{8.60}$	&$\mathbf{0.93}$	\\
\object{120}	&\object{GDDS$\,$12-7939}	&$8.11\pm0.10$	&$2.00$	&$0.33$	&$\mathbf{8.45\pm0.19}$	&$\mathbf{4.09}$	&$\mathbf{0.67}$	\\
\object{122}	&\object{GDDS$\,$22-0040}	&$8.04\pm0.15$	&$2.11$	&$0.43$	&$\mathbf{8.55\pm0.25}$	&$\mathbf{2.76}$	&$\mathbf{0.57}$	\\
\object{123}	&\object{GDDS$\,$22-0145}	&$8.10\pm0.23$	&$2.10$	&$0.43$	&$\mathbf{8.72\pm0.19}$	&$\mathbf{2.73}$	&$\mathbf{0.57}$	\\
\object{124}	&\object{GDDS$\,$22-0563}	&$7.99\pm0.17$	&$2.52$	&$0.43$	&$\mathbf{8.78\pm0.16}$	&$\mathbf{3.30}$	&$\mathbf{0.57}$	\\
\object{125}	&\object{GDDS$\,$22-0619}	&$8.19\pm0.26$	&$2.00$	&$0.39$	&$\mathbf{8.73\pm0.16}$	&$\mathbf{3.10}$	&$\mathbf{0.61}$	\\
\object{126}	&\object{GDDS$\,$22-0630}	&..	&..	&..	&$\mathbf{9.20\pm0.11}$	&$\mathbf{9.32}$	&$\mathbf{1.00}$	\\
\object{127}	&\object{GDDS$\,$22-0643}	&$8.28\pm0.24$	&$2.78$	&$0.50$	&$\mathbf{8.68\pm0.12}$	&$\mathbf{2.83}$	&$\mathbf{0.50}$	\\
\object{128}	&\object{GDDS$\,$22-0751}	&$8.30\pm0.25$	&$3.06$	&$0.50$	&$\mathbf{8.73\pm0.07}$	&$\mathbf{3.07}$	&$\mathbf{0.50}$	\\
\object{129}	&\object{GDDS$\,$22-0926}	&$8.07\pm0.15$	&$1.74$	&$0.37$	&$\mathbf{8.51\pm0.24}$	&$\mathbf{3.02}$	&$\mathbf{0.63}$	\\
\object{131}	&\object{GDDS$\,$22-1674}	&$8.09\pm0.12$	&$2.23$	&$0.42$	&$\mathbf{8.51\pm0.23}$	&$\mathbf{3.09}$	&$\mathbf{0.58}$	\\
\object{132}	&\object{GDDS$\,$22-2196}	&$8.12\pm0.10$	&$1.50$	&$0.25$	&$\mathbf{8.43\pm0.19}$	&$\mathbf{4.48}$	&$\mathbf{0.75}$	\\
\object{133}	&\object{GDDS$\,$22-2491}	&$\mathbf{8.12\pm0.13}$	&$\mathbf{4.48}$	&$\mathbf{0.65}$	&$8.53\pm0.18$	&$2.36$	&$0.35$	\\
\object{134}	&\object{GDDS$\,$22-2541}	&$8.10\pm0.24$	&$2.21$	&$0.43$	&$\mathbf{8.78\pm0.16}$	&$\mathbf{2.91}$	&$\mathbf{0.57}$	\\
\object{135}	&\object{GDDS$\,$22-2639}	&$7.76\pm0.03$	&$1.81$	&$0.20$	&$\mathbf{9.08\pm0.12}$	&$\mathbf{7.33}$	&$\mathbf{0.80}$	\\
\object{136}	&\object{SKK2001 368}	&..	&..	&..	&$\mathbf{9.12\pm0.06}$	&$\mathbf{15.57}$	&$\mathbf{1.00}$	\\
\object{137}	&\object{SKK2001 159}	&$\mathbf{8.65\pm0.10}$	&$\mathbf{10.06}$	&$\mathbf{1.00}$	&$9.42\pm0.01$	&$0.02$	&$0.00$	\\
\object{138}	&	&$\mathbf{8.28\pm0.27}$	&$\mathbf{2.81}$	&$\mathbf{0.54}$	&$8.74\pm0.10$	&$2.41$	&$0.46$	\\
\object{140}$^*$	&	&$\mathbf{8.29\pm0.20}$	&$\mathbf{3.48}$	&$\mathbf{0.64}$	&$8.60\pm0.15$	&$1.94$	&$0.36$	\\
\object{141}	&	&$7.90\pm0.07$	&$1.41$	&$0.08$	&$\mathbf{8.69\pm0.06}$	&$\mathbf{15.32}$	&$\mathbf{0.92}$	\\
\end{longtable}}

\twocolumn

\appendix

\section{Discussion on the fitting method\label{sec:fitmethod}}

\begin{figure}
\begin{center}\includegraphics[%
  width=1.0\columnwidth,
  keepaspectratio]{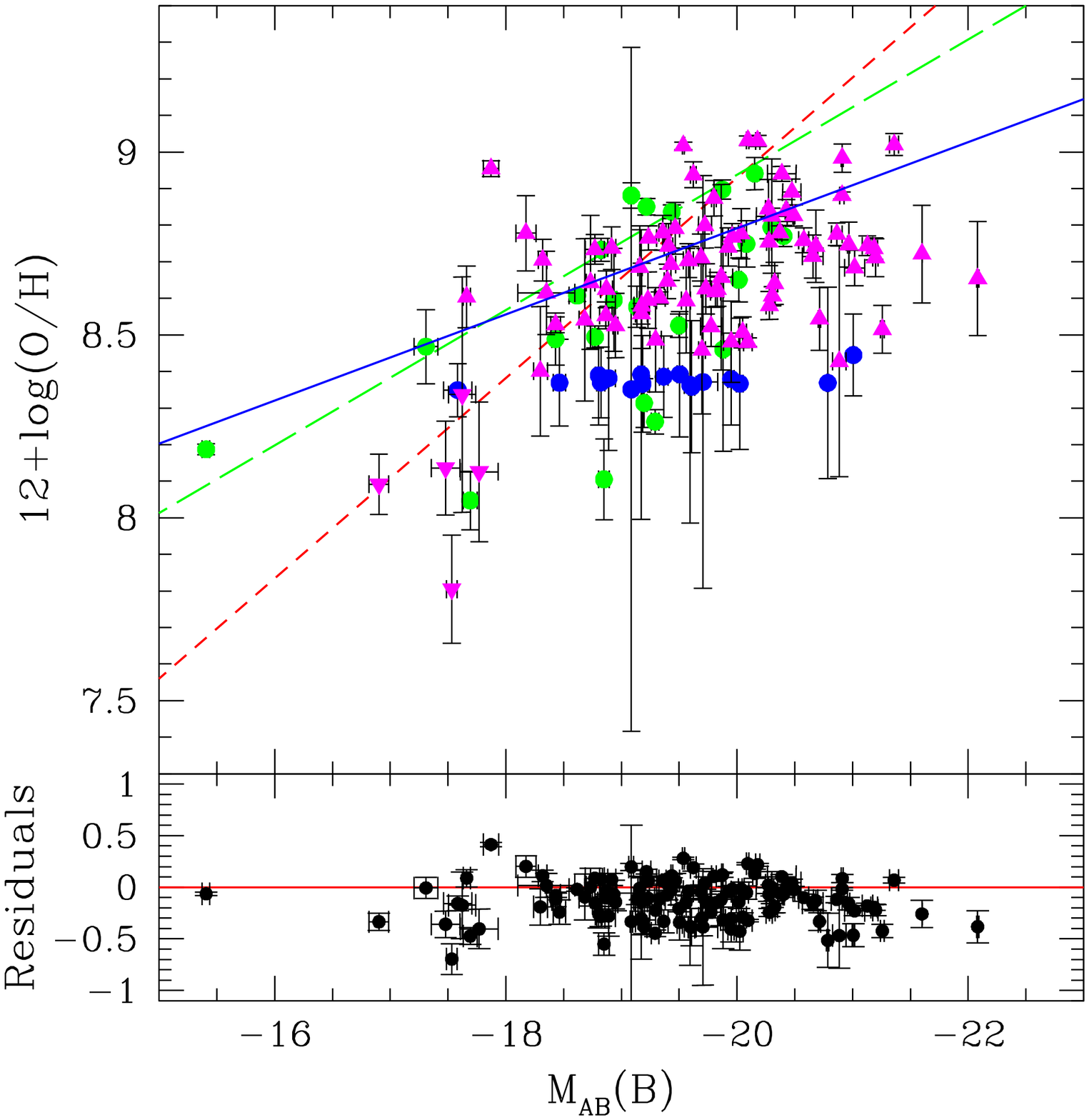}\end{center}

\caption{The Luminosity-Metallicity relation in the $B$ band with the \emph{x-and-y}
method. Same legend as in Fig.~\ref{ohb}.}

\label{ohlsig}
\end{figure}

The linear regression used to derive the $L-Z$ relation is based
on the OLS linear bisector method \citep{Isobe:1990ApJ...364..104I},
it gives the bisector of the two least-squares regressions \emph{x-on-y}
and \emph{y-on-x}. This method is useful when we know that the errors
in the two variables $x$ and $y$ are independent, without knowing
the exact values of these errors (e.g. this was the case with the
2dFGRS data). For our sample of intermediate-redshift galaxies, we
have an estimate of the errors. We thus try to take advantage of these
errors to do the best mathematical fit of our data.

Fig.~\ref{ohlsig} shows the results of the maximum likelihood fit
weighted with errors in the two variables (the \emph{x-and-y} method).
We find the following relation:\begin{equation}
12+\log(\textrm{O/H})=-0.12\cdot M_{AB}(B)+6.44\label{eq:lzsig}\end{equation}
 with a rms of the residuals of $0.20$ dex. This relation is very
close to the local relation derived by \citet{Tremonti:2004astro.ph..5537T}
with the SDSS data, but has significant differences with the one derived
by \citet{Lamareille:2004MNRAS.350..396L} with the 2dFGRS data. We
highlight here the \emph{high sensitivity of the $L-Z$ relation to
the fitting method}.

The \emph{x-and-y} method suffers from an important drawback: it is
very sensitive to the ratio between the errors in the two variables.
Indeed, the \emph{x-and-y} method actually does an average of the
\emph{y-on-x} and \emph{x-on-y} fits, using the ratio between the
errors as a weight (e.g. in our data the errors in the luminosity
are almost negligible compared to the errors in the metallicity, so
that the \emph{x-and-y} method tends to be equivalent the the \emph{x-on-y}
fit).

In our sample, we cannot compare in an absolute way the errors in
the luminosity to the ones in the metallicity. Taking this into account,
we decided to keep the OLS bisector method for further studies, even
if the best way to derive the $L-Z$ relation is to do an extensive
study of all sources of errors and to use the \emph{x-and-y} method.
\end{document}